\documentclass[aip,reprint,amsmath,amssymb,superscriptaddress]{revtex4-2}

\usepackage{verbatim} 

\usepackage{graphicx}
\usepackage{dcolumn}

\usepackage{amstext,amsmath,amssymb,amsfonts}
\usepackage{bm}

\usepackage{txfonts}
\usepackage{mathtools}
\DeclareMathAlphabet{\mathpzc}{OT1}{pzc}{m}{it}
\DeclareFontFamily{OT1}{pzc}{}
\DeclareFontShape{OT1}{pzc}{m}{it}{<-> s * [1.100] pzcmi7t}{}
\DeclareMathAlphabet{\mathpzc}{OT1}{pzc}{m}{it}

\usepackage{hyperref}

\usepackage{color}
\definecolor{lightblue}{rgb}{0.2,0.2,0.7}
\definecolor{darkblue}{rgb}{0,0.25,0.5}
\definecolor{redbrown}{rgb}{0.875,0.25,0.125}
\definecolor{darkgreen}{rgb}{0,0.5,0}

\newcommand{\braket}[2]{\ensuremath{\langle  #1, #2  \rangle}}

\renewcommand{\H}{\ensuremath{\text{H}}}

\renewcommand{\l}{\ensuremath{\lambda}}
\newcommand{\w}{\ensuremath{\text{w}}}

\newcommand{\ee}{\ensuremath{\text{ee}}}
\renewcommand{\ne}{\ensuremath{\text{ne}}}

\newcommand{\HF}{\ensuremath{\text{HF}}}

\renewcommand{\d}{\ensuremath{\text{d}}}
\newcommand{\s}{\ensuremath{\text{s}}}

\newcommand{\x}{\ensuremath{\text{x}}}

\renewcommand{\c}{\ensuremath{\text{c}}}
\newcommand{\Hxc}{\ensuremath{\text{Hxc}}}
\newcommand{\Hx}{\ensuremath{\text{Hx}}}
\newcommand{\md}{\ensuremath{\text{md}}}

\newcommand{\argmax}[1]{\underset{#1}{\text{argmax}}}

\newcommand{\isEquivTo}[1]{\underset{#1}{\sim}}

\renewcommand{\i}{\ensuremath{\text{i}}}
\newcommand{\B}{\ensuremath{{\cal B}}}
\newcommand{\FCI}{\ensuremath{\text{FCI}}}

\newcommand{\calh}{\ensuremath{\mathpzc{h}}}

\begin{document}

\title{Basis-set correction based on density-functional theory: Rigorous framework for a one-dimensional model}

\author{Diata Traore}
\email{dtraore@lct.jussieu.fr}
\affiliation{Laboratoire de Chimie Théorique, Sorbonne Université and CNRS, F-75005 Paris, France}

\author{Emmanuel Giner}
\email{emmanuel.giner@lct.jussieu.fr}
\affiliation{Laboratoire de Chimie Théorique, Sorbonne Université and CNRS, F-75005 Paris, France}

\author{Julien Toulouse}
\email{toulouse@lct.jussieu.fr}
\affiliation{Laboratoire de Chimie Théorique, Sorbonne Université and CNRS, F-75005 Paris, France}
\affiliation{Institut Universitaire de France, F-75005 Paris, France}

\date{14 December, 2021}

\begin{abstract}
We reexamine the recently introduced basis-set correction theory based on density-functional theory consisting in correcting the basis-set incompleteness error of wave-function methods using a density functional. We use a one-dimensional model Hamiltonian with delta-potential interactions which has the advantage of making easier to perform a more systematic analysis than for three-dimensional Coulombic systems while keeping the essence of the slow basis convergence problem of wave-function methods. We provide some mathematical details about the theory and propose a new variant of basis-set correction which has the advantage of being suited to the development of an adapted local-density approximation. We show indeed how to develop a local-density approximation for the basis-set correction functional which is automatically adapted to the basis set employed, without resorting to range-separated density-functional theory as in previous works, but using instead a finite uniform electron gas whose electron-electron interaction is projected on the basis set. The work puts the basis-set correction theory on firmer grounds and provides an interesting strategy for the improvement of this approach.
\end{abstract}

\maketitle

\section{Introduction}

In electronic-structure theory of atoms, molecules or solids, one of the main limitations of standard correlated wave-function computational methods for solving the Schr\"odinger equation is the slow convergence of the energy and other properties with respect to the size of the one-electron basis set employed (see, e.g., Refs.~\onlinecite{HelKloKocNog-JCP-97,HalHelJorKloKocOlsWil-CPL-98,HelJorOls-BOOK-02,SheGruBooKreAla-PRB-12}). This slow basis convergence originates from the singular behavior of the repulsive Coulomb electron-electron interaction at small interelectronic distances, which creates a depletion in the wave function with a characteristic derivative discontinuity at electron-electron coalescence --- the infamous electron-electron cusp~\cite{Kat-CPAM-57,PacBye-JCP-66}. 

The two main approaches for dealing with this problem are (i) extrapolation to the complete-basis-set limit by using increasingly large basis sets~\cite{HelKloKocNog-JCP-97,HalHelJorKloKocOlsWil-CPL-98}, and (ii) explicitly correlated methods which incorporate in the wave function a correlation factor reproducing the electron-electron cusp (see, e.g., Ref.~\onlinecite{HatKloKohTew-CR-12}). Recently, some of the present authors introduced an alternative basis-set correction scheme based on density-functional theory (DFT)~\cite{GinPraFerAssSavTou-JCP-18}. This latter scheme consists in correcting the energy calculated by a wave-function method with a finite basis set by a density functional incorporating the short-range electron correlation effects missing in the basis set. This basis-set correction scheme was further developed and tested in Refs.~\onlinecite{LooPraSceTouGin-JPCL-19,GinSceTouLoo-JCP-19,LooPraSceGinTou-JCTC-20,GinSceLooTou-JCP-20,YaoGinLiTouUmr-JCP-20,GinTraPraTou-JCP-21,YaoGinAndTouUmr-JCP-21}, demonstrating that it successfully accelerates the basis convergence of wave-function methods for various properties and systems.

The advantages of the basis-set correction scheme is its conceptual simplicity and computational efficiency. In practice, however, the limpidity of this approach is somewhat diminished by the fact that in all previously cited works the basis-set correction functional was approximated by short-range correlation functionals borrowed from range-separated DFT~\cite{Sav-INC-96,TouColSav-PRA-04,TouGorSav-TCA-05,PazMorGorBac-PRB-06,FerGinTou-JCP-19}), relying on an approximate mapping between the basis-set correction theory and range-separated DFT.

In the present work, we reexamine more closely the basis-set correction theory. For this, we use a one-dimensional (1D) model Hamiltonian with delta-potential interactions~\cite{Ros-JCP-71,Lap-AJP-75,HerSti-PRA-75,MagBur-PRA-04} which has the advantage of making easier to perform a more systematic analysis than for three-dimensional (3D) Coulombic systems, while keeping the essence of the slow basis convergence problem of wave-function methods. After introducing the 1D model and discussing its relevance in Section~\ref{sec:1Dmodel}, we present the basis-set correction theory in some mathematical details in Section~\ref{sec:theory}. In particular, we introduce a new variant of basis-set correction which has the advantage of being suited for development of an adapted local-density approximation (LDA). In Section~\ref{sec:LDA}, we show indeed how to develop a LDA for the basis-set correction functional which is automatically adapted to the basis set employed without resorting to range-separated DFT but using instead a finite uniform electron gas (UEG) whose electron-electron interaction is projected on the basis set. Section~\ref{sec:conclusion} contains our conclusion and outlook. Hartree atomic units (a.u.) are used throughout this work.

\section{One-dimensional model system}
\label{sec:1Dmodel}

\subsection{Description of the model}

We consider $N=2$ electrons in a 1D He-like atom with delta-potential interactions described by the Hamiltonian~\cite{Ros-JCP-71,Lap-AJP-75,HerSti-PRA-75,MagBur-PRA-04}
\begin{eqnarray}
H = T + W_\ee + V_\ne,
\label{H}
\end{eqnarray}
where
\begin{eqnarray}
T = -\frac{1}{2} \sum_{i=1}^N \frac{\partial^2}{\partial x_i^2},\;
W_\ee = \delta(x_1-x_2), \;
V_\ne = -Z \sum_{i=1}^N \delta(x_i) \;\;
\label{TWV}
\end{eqnarray}
are the kinetic-energy operator, the Dirac-delta electron-electron interaction, and the Dirac-delta nucleus-electron potential with nuclear charge $Z=2$, respectively. Since we will be only interested in the spin-singlet ground state, we can ignore spin and antisymmetry, and thus work on the one-electron Hilbert space $\calh = L^2(\mathbb{R}, \mathbb{C})$ and the two-electron (non-antisymmetrized) tensor-product Hilbert space ${\cal H} = \calh \otimes \calh$. The ground-state energy can be expressed as 
\begin{eqnarray}
E_0 = \min_{\Psi \in {\cal W}} \braket{\Psi}{H\Psi},
\label{E0}
\end{eqnarray}
where ${\cal W}$ is the set of all admissible wave functions
\begin{eqnarray}
{\cal W} = \Big\{ \Psi \in {\cal H}\;|\; \Psi \in H^1(\mathbb{R}^2, \mathbb{C}),\; \braket{\Psi}{\Psi}=1 \Big\},
\label{calW}
\end{eqnarray}
where $H^1(\mathbb{R}^2, \mathbb{C}) =\{ f \in L^2(\mathbb{R}^2, \mathbb{C}) \;|\; \partial_i f \in L^2(\mathbb{R}^2, \mathbb{C}), i=1,2\}$ is the first-order Sobolev space and $\braket{\;}{}$ designates the $L^2$ inner product.

The ground state of this 1D He-like atom with delta-potential interactions can be considered as a model for the ground state of the real 3D He atom. Indeed, it can be shown that the ground state of a generalization to arbitrary dimension $D$ of the electronic Hamiltonian of the He atom with Coulomb-potential interactions exactly reduces for $D=1$, after appropriate scaling of the energies and distances, to the ground state of the Hamiltonian in Eq.~(\ref{H})~\cite{HerSti-PRA-75,Her-JCP-86,DorHer-PRA-86,DorHer-JCP-87,HerAveGos-BOOK-93}. Our main interest in this model is that it gives an electron-electron cusp (or derivative discontinuity) condition identical to the familiar 3D one~\cite{Kat-CPAM-57,PacBye-JCP-66}, i.e. for small interelectronic distances $x_{12}=x_{1}-x_{2}$ the exact ground-state wave function behaves as
\begin{eqnarray}
\Psi_0(x_1,x_2) = \Psi_0(x_1,x_1) \left( 1 + \frac{1}{2} |x_{12}| + O(x_{12}^2) \right).
\label{eecusp}
\end{eqnarray}
When using a finite one-electron basis set, we thus expect a slow convergence with the basis size very similar to the slow convergence observed in 3D quantum systems with the Coulomb electron-electron interaction. This is why we prefer this model to other 1D quantum systems (see, e.g., Refs.~\onlinecite{LooBalGil-PCCP-15,BalLooGil-PCCP-17}). 

Another neat fact about the present model is that it can be solved analytically at the Hartree-Fock (HF) level~\cite{NogValDij-AJP-76}. The total HF ground-state energy is
\begin{eqnarray}
E_\HF  = -Z^2 + \frac{Z}{2} - \frac{1}{12} = -3.083333... \; \text{a.u.},
\label{EHF}
\end{eqnarray}
and the HF (doubly) occupied spatial orbital is
\begin{eqnarray}
\forall x \in \mathbb{R},\; \phi_1(x) =  2 \beta \sqrt{\gamma} \frac{e^{-\beta |x|}}{1-\gamma\; e^{-2\beta |x|}},
\label{phi1}
\end{eqnarray}
with $\beta=Z-1/2=3/2$ and $\gamma=1/(4Z-1) = 1/7$. The exact ground-state energy cannot be calculated analytically but has been accurately estimated numerically~\cite{Ros-JCP-71,MagBur-PRA-04} to be $E_0 = -3.155390$ a.u..

\subsection{Full-configuration interaction in a basis set}
\label{sec:FCIB}

We now consider full-configuration-interaction (FCI) calculations in a finite one-electron basis set $\B \subset H^1(\mathbb{R},\mathbb{C})$. To have a systematically improvable basis set, we use Hermite (or Hermite-Gaussian) basis functions
\begin{eqnarray}
\forall x \in \mathbb{R},\; f_n^\alpha(x)= N_n^\alpha \; H_n(\sqrt{2\alpha}x) \; e^{-\alpha x^2}, 
\label{fnx}
\end{eqnarray}
where $n$ is a natural number, $H_n$ are the Hermite polynomials, $N_n^\alpha = (2^n n!)^{-1/2} (2\alpha/\pi)^{1/4}$ is the normalization factor, and $\alpha >0$ is a real constant. It is well known that the set $\{ f_n^\alpha\}_{n=0,...,n_\text{max}}$ converges to a complete orthonormal basis of $L^2(\mathbb{R}, \mathbb{C})$ in the limit $n_\text{max}\to \infty$ for any fixed exponent $\alpha$. We deliberately use the same exponent in all basis functions, namely $\alpha=11.5$, instead of multiple exponents in order to avoid optimizing them. Except for that, this basis set is quite similar to the Gaussian-type-orbital basis sets widely used in quantum chemistry. Since we are not interested in the convergence of the HF energy with this basis set (which is slow, see Appendix~\ref{app:1e}) but only in the convergence of the FCI correlation energy, we add the exact occupied HF orbital $\phi_1$ given in Eq.~(\ref{phi1}) to the basis set. Our final basis set is thus
\begin{eqnarray}
\B = \Big\{ \phi_1 \Big\} \cup \Big\{ f_n^\alpha \Big\}_{n=0,...,n_\text{max}} \equiv \Big\{ \chi_\mu \Big\}_{\mu=1,...,M},
\label{}
\end{eqnarray}
and contains $M=n_\text{max}+2$ basis functions: $\chi_1=\phi_1$, $\chi_2=f_0^\alpha$, ..., $\chi_M=f_{n_\text{max}}^\alpha$.

We introduce now $\calh^\B=\text{span}(\B)$ as the $M$-dimensional one-electron Hilbert space spanned by this basis set $\B$, and ${\cal H}^{\B} = \calh^\B \otimes \calh^\B$ the corresponding two-electron Hilbert space of dimension $M^2$. The FCI ground-state energy for this basis set $\B$ is
\begin{eqnarray}
E^{\B}_{\FCI} = \min_{\Psi \in {\cal W}^{\B}} \braket{\Psi}{H \Psi},
\label{E0BFCI}
\end{eqnarray}
where ${\cal W}^\B$ is the set of normalized wave functions restricted to ${\cal H}^{\B}$
\begin{eqnarray}
{\cal W}^\B = \Big\{ \Psi \in {\cal H}^\B \;|\; \braket{\Psi}{\Psi}=1 \Big\}.
\label{calWB}
\end{eqnarray}
In practice, we proceed as follows. For each basis size $n_\text{max}$, we first perform a HF calculation~\cite{SzaOst-BOOK-96,HelJorOls-BOOK-02}. The nucleus-electron integrals are just $\braket{\chi_\mu}{v_{\ne} \chi_\nu}=-Z\chi_\mu(0)\chi_\nu(0)$, and the kinetic integrals $\braket{\chi_\mu}{t \chi_\nu}=(1/2)\int_\mathbb{R} \chi_\mu'(x)\chi_\nu'(x)\d x$ and the two-electron integrals $\braket{\chi_\mu\chi_\nu}{W_\ee \chi_\l\chi_\sigma} = \int_\mathbb{R} \chi_\mu(x)\chi_\nu(x) \chi_\l(x)\chi_\sigma(x) \d x$ are calculated by Romberg numerical integration~\cite{PreTeuVetFla-BOOK-92}. We then use the obtained HF orbitals $\{ \phi_i \}_{i=1,...,M}$ to expand the FCI ground-state wave function as
\begin{eqnarray}
\Psi_{\FCI}^\B(x_1,x_2) = \sum_{i=1}^M \sum_{i=1}^M c_{i,j} \phi_i(x_1) \phi_j(x_2).
\end{eqnarray}
The FCI coefficients $c_{i,j}$ and the associated FCI ground-state energy $E_{\FCI}^\B$ are found by diagonalization of the Hamiltonian. Parity inversion symmetry is exploited in all our calculations.

\begin{figure}
\includegraphics[scale=0.35,angle=-90]{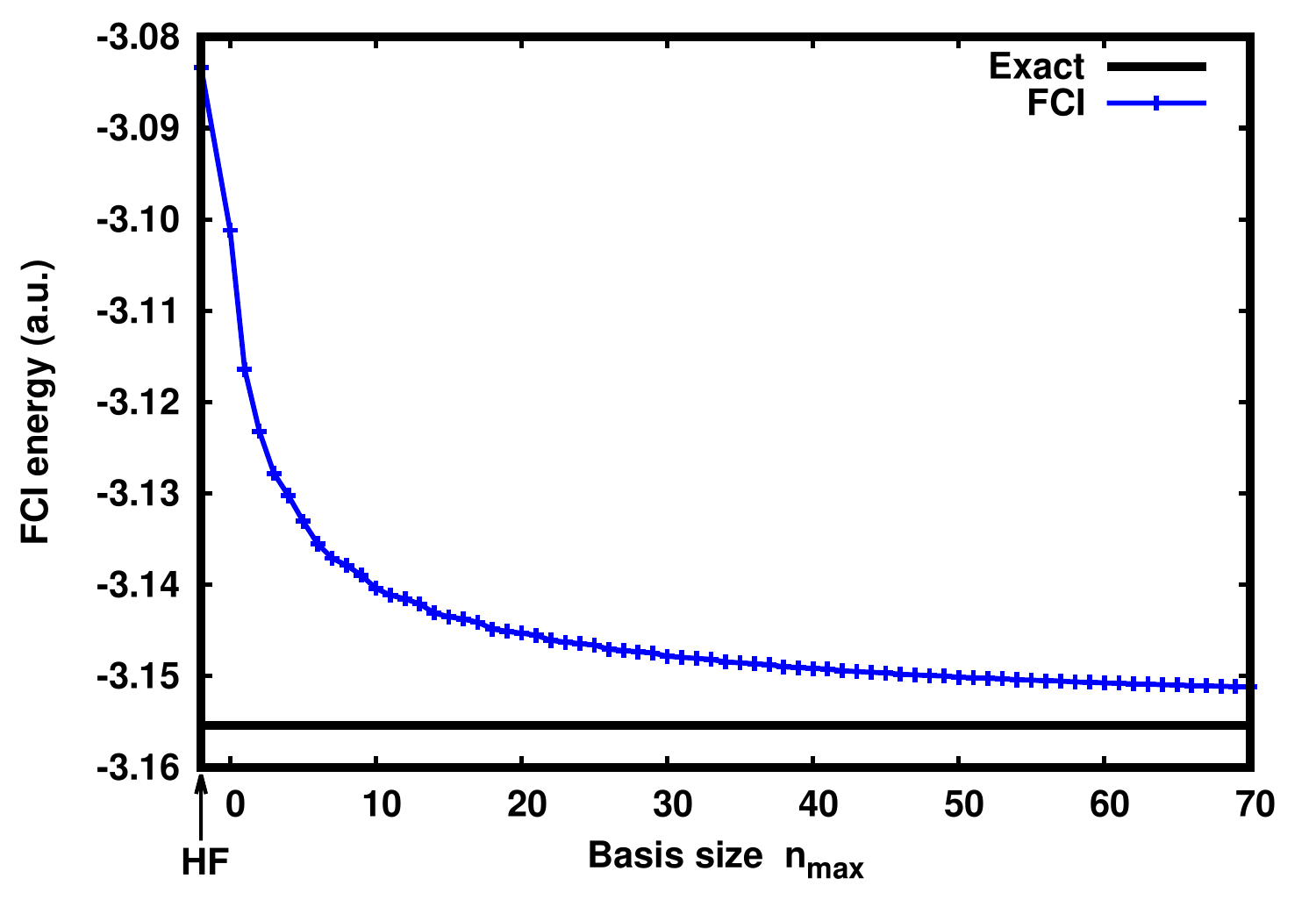}
\caption{FCI ground-state energy $E_{\FCI}^\B$ [Eq.~(\ref{E0BFCI})] of the 1D He-like atom as a function of the basis size $n_\text{max}$. The exact energy is taken from Ref.~\onlinecite{MagBur-PRA-04}.}
\label{fig:FCIenergy}
\end{figure}

\begin{figure}
\includegraphics[scale=0.35,angle=-90]{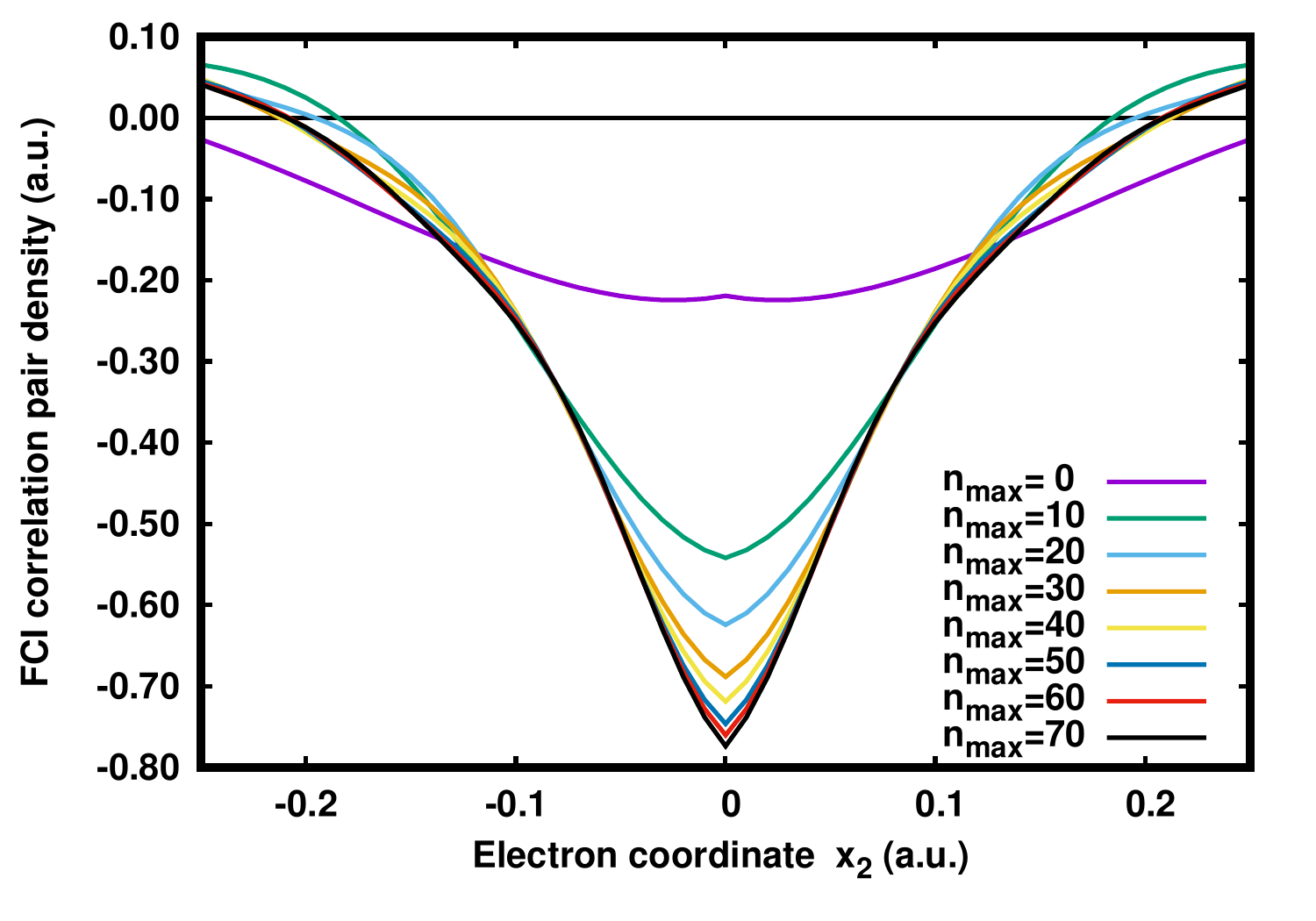}
\caption{FCI correlation pair density $\rho_{2,\c}^\B(x_1,x_2)$ of the 1D He-like atom at $x_1=0$ as a function of $x_2$ for different basis size $n_\text{max}$.}
\label{fig:FCIpairdensity}
\end{figure}

In Fig.~\ref{fig:FCIenergy}, we report the FCI ground-state energy $E_{\FCI}^\B$ as a function of the basis size $n_\text{max}$. We observe a quite slow convergence of $E_{\FCI}^\B$ with $n_\text{max}$ toward the exact ground-state energy $E_0$. A numerical fit from $n_\text{max}=50$ and $70$ gives the following power-law convergence
\begin{eqnarray}
E_{\FCI}^\B \isEquivTo{n_{\text{max}} \to \infty} E_0 + \frac{A}{n_{\text{max}}^b},
\end{eqnarray}
gives $A\approx 0.077$ a.u. and $b \approx 0.68$. Equivalently, in terms of the number of basis functions $M$, this convergence law can be stated as $1/M^b$. Note that since our basis set includes the exact HF occupied orbital this slow convergence is entirely due to the correlation energy. According to the analysis given in Appendix~\ref{app:2e}, we theoretically expect $b=0.5$. The difference most likely means that larger values of $n_\text{max}$ are needed to reach the asymptotic regime. Note that for the 3D Coulomb case it is well known that the correlation energy exhibits a cubic-law convergence with respect to either the maximal angular momentum or the maximal principal quantum number of the basis set, or equivalently a $1/M$ convergence law in terms of the number of basis functions~\cite{KloBakJorOlsHel-JPB-99,HelJorOls-BOOK-02}. In the present work, the use of a basis of Hermite functions with a single exponent thus leads to an even slower convergence rate.

To illustrate further the slow basis convergence, we also calculate the FCI correlation pair density
\begin{eqnarray}
\rho_{2,\c}^\B(x_1,x_2) = 2\left(|\Psi_{\FCI}^\B(x_1,x_2)|^2 - |\Phi_\HF(x_1,x_2)|^2\right),
\end{eqnarray}
where $\Phi_\HF(x_1,x_2)=\phi_1(x_1)\phi_1(x_2)$ is the HF wave function. In Fig.~\ref{fig:FCIpairdensity}, this quantity is plotted with respect to the second electron coordinate $x_2$ for a fixed value of the first electron coordinate $x_1=0$. The convergence of $\rho_{2,\c}^\B(x_1,x_2)$ with $n_\text{max}$ is again slow and reminiscent of the well-known slow basis convergence of the correlation pair density for the 3D Coulomb electron-electron interaction (see, e.g., Refs.~\onlinecite{HelJorOls-BOOK-02,TewKloHel-JCC-07,FraMusLupTou-JCP-15}). Note that the small derivative discontinuity seen on all the curves of Fig.~\ref{fig:FCIpairdensity} at $x_2=0$ is due to the fact that we include in our basis set the exact HF orbital in Eq.~(\ref{phi1}) which has itself an electron-nucleus cusp, namely $\phi_1(x) = \phi_1(0) (1 -Z |x| + O(x^2))$. The electron-electron cusp condition in Eq.~(\ref{eecusp}) is only recovered for large $n_\text{max}$.

In conclusion, the present 1D model adequately captures the main characteristics of the slow basis convergence problem of standard quantum-chemistry wave-function methods, and it is thus appropriate for applying our basis-set correction approach.

\section{Basis-set correction theory}
\label{sec:theory}

We now develop the basis-set correction theory based on DFT for the present 1D model which aims at removing the basis-set incompleteness error in the FCI ground-state energy. This requires to develop an extension of standard DFT from the usual complete-basis-set setting to the case of the incomplete finite one-electron basis set $\B$. In such a finite basis set, it is known that the original Hohenberg-Kohn theorem~\cite{HohKoh-PR-64} does not hold anymore~\cite{Har-PRA-83,GorErn-PRA-95,PinBokLudBoa-TCA-09}, in the sense that there is an infinite number of local potentials which give, after projection in a finite basis set, the same ground-state density~\cite{StaScuDav-JCP-06}. However, we will show that we can still define density functionals associated with a finite basis set.

\subsection{Density-functional theory for the one-dimensional model}
\label{sec:standardDFT}

We start by reviewing some useful definitions of standard DFT specialized to the 1D model. For mathematically oriented reviews of DFT, see for instance Refs.~\onlinecite{Lie-IJQC-83,DreGro-BOOK-90,Esc-BOOK-03,AnaCan-AN-09,EngDre-BOOK-11,KvaEksTeaHel-JCP-14,Hel-TALK-17,Can-TALK-19}, and in particular Ref.~\onlinecite{LewLieSei-ARX-20} which encompasses the 1D case.

Working on the same Hilbert space ${\cal H}$ as before, we now consider the following 1D Hamiltonian for $N=2$ electrons with a general external potential $v$
\begin{eqnarray}
H[v] = T + W_\ee + V,
\label{Hv}
\end{eqnarray}
where again $T = -(1/2) \sum_{i=1}^N \partial^2/\partial x_i^2$ and $W_\ee = \delta(x_1-x_2)$, and $V = \sum_{i=1}^N v(x_i)$ is now a general external potential operator. We will still take admissible wave functions to be in the space ${\cal W}$ given in Eq.~(\ref{calW}). The convex set of densities representable by a wave function $\Psi \in {\cal W}$, the so-called $N$-representable densities, is then~\cite{Lie-IJQC-83,LewLieSei-ARX-20}
\begin{eqnarray}
{\cal R} &=& \left\{ \rho \; | \; \exists \; \Psi \in {\cal W}, \rho_\Psi  = \rho \right\} 
\nonumber\\
&=& \left\{\rho \in L^1(\mathbb{R}) \;|\; \rho \geq 0, \; \int_{\mathbb{R}} \rho(x) \d x = N, \; \sqrt{\rho} \in H^1(\mathbb{R}) \right\},
\nonumber\\
\label{RN}
\end{eqnarray}
where $\rho_\Psi(x_1) = N \int_{\mathbb{R}} |\Psi(x_1,x_2)|^2 \d x_2$ is the density of the wave function $\Psi$. We have ${\cal R} \subset {\cal X}$ where ${\cal X}$ is the Banach space ${\cal X} = C_0(\mathbb{R}) \cap L^1(\mathbb{R})$ with $C_0(\mathbb{R})$ the space of continuous functions vanishing at infinity. Therefore, the space of external potentials $v$ that we can consider is the continuous dual space of ${\cal X}$, i.e. ${\cal V}= {\cal X}' = M(\mathbb{R}) + L^\infty(\mathbb{R})$ where $M(\mathbb{R})$ is the space of bounded Radon measures. Note that the set ${\cal V}$ includes the external potential considered in Section~\ref{sec:1Dmodel}, i.e. $v_\ne(x)=-Z\delta(x)$.
For $v\in {\cal V}$, we then define the ground-state energy as
\begin{eqnarray}
E_0[v] = \inf_{\Psi \in {\cal W}} \braket{\Psi}{H[v]\Psi}.
\label{E0v}
\end{eqnarray}

The Levy-Lieb density functional~\cite{Lev-PNAS-79,Lie-IJQC-83} is defined as a constrained-search over wave functions yielding the density $\rho$
\begin{eqnarray}
\forall \rho \in {\cal R}, \; F[\rho] = \min_{\Psi \in {\cal W}_\rho} \braket{\Psi}{(T + W_\ee)\Psi},
\label{FLLrho}
\end{eqnarray}
where ${\cal W}_\rho = \left\{ \Psi \in {\cal W}\;|\; \rho_{\Psi} = \rho \right\}$. It gives the ground-state energy as
\begin{eqnarray}
E_0[v] = \inf_{\rho \in {\cal R}} \left( F[\rho] + (v,\rho) \right),
\label{E0vrhoLLRN}
\end{eqnarray}
where we have introduced the notation $(v,\rho) = \int_{\mathbb{R}} v(x) \rho(x) \d x$. If a minimizing density $\rho_0$ exists in Eq.~(\ref{E0vrhoLLRN}) then it is an exact ground-state density for the potential $v$.

One can also define the Lieb density functional~\cite{Lie-IJQC-83}, which is the Legendre–Fenchel convex-conjugate of $E_0[v]$
\begin{eqnarray}
\forall \rho \in {\cal R},\; F_\text{L}[\rho] = \sup_{v \in {\cal V}} \left( E_0[v] - (v,\rho) \right).
\label{FLrho}
\end{eqnarray}
Just like the Levy-Lieb functional $F$, the Lieb functional $F_\text{L}$ gives the exact ground-state energy as $E_0[v] = \inf_{\rho \in {\cal R}} \left( F_\text{L}[\rho] + (v,\rho) \right)$. In general, the functionals $F$ and $F_\text{L}$ are different, the Lieb functional being in fact the lower semi-continuous convex envelope (lscv) of the Levy-Lieb functional, i.e.
\begin{eqnarray}
F_\text{L} = \text{lscv}(F) \leq F.
\label{FDMLLL}
\end{eqnarray}
It turns out that the Lieb functional can also be expressed as a generalization of the Levy-Lieb functional in which the constrained search is extended from pure-state wave functions to ensemble density matrices~\cite{Val-JCP-80a,Lie-IJQC-83}. This implies that $F[\rho] = F_\text{L}[\rho]$ for densities $\rho$ which are densities of a \textit{non-degenerate} ground state of the Hamiltonian $H[v]$ for some potential $v$. In the present case of two spin-singlet electrons, the ground state is always non-degenerate and thus the Levy-Lieb and Lieb functionals are identical, i.e. $F=F_\text{L}$.

\subsection{First variant of basis-set correction}

The first variant of basis-set correction corresponds to the one introduced for the 3D Coulombic case in Ref.~\onlinecite{GinPraFerAssSavTou-JCP-18} and further developed in Refs.~\onlinecite{LooPraSceTouGin-JPCL-19,GinSceTouLoo-JCP-19,LooPraSceGinTou-JCTC-20,GinSceLooTou-JCP-20,GinTraPraTou-JCP-21}. We consider the Hamiltonian $H[v]$ in Eq.~(\ref{Hv}) on the two-electron Hilbert space ${\cal H}^B$ associated with the basis set $\B$. For $v\in{\cal V}$, the FCI ground-state energy is
\begin{eqnarray}
E_{\FCI}^{\B}[v] = \min_{\Psi \in {\cal W}^{\B}} \braket{\Psi}{H[v]\Psi},
\label{EFCIBv}
\end{eqnarray}
where ${\cal W}^\B$, given in Eq.~(\ref{calWB}), is the set of normalized wave functions restricted to ${\cal H}^{\B}$. We define the corresponding Levy-Lieb density functional for the basis set $\B$ as
\begin{eqnarray}
\forall \rho \in {\cal R}^{\B}, \; F^{\B}[\rho] = \min_{\Psi \in {\cal W}^{\B}_\rho} \braket{\Psi}{(T + W_\ee)\Psi},
\label{FLLBrho}
\end{eqnarray}
where ${\cal W}^\B_\rho = \left\{ \Psi \in {\cal W}^\B\;|\; \rho_{\Psi} = \rho \right\}$ and ${\cal R}^{\B}$ is the set of densities representable by a wave function $\Psi \in {\cal W}^{\B}$
\begin{eqnarray}
{\cal R}^{\B} &=& \left\{ \rho \; | \; \exists \; \Psi \in {\cal W}^{\B}, \rho_\Psi  = \rho \right\}.
\end{eqnarray}
A priori, this set is not convex and not easily characterized. The FCI ground-state energy can be expressed as
\begin{eqnarray}
E_{\FCI}^{\B}[v] = \min_{\rho \in {\cal R}^{\B}} \left( F^{\B}[\rho] + (v,\rho) \right).
\label{E0B1vrhoLLRN}
\end{eqnarray}

We now decompose the exact Levy-Lieb density functional $F[\rho]$ in Eq.~(\ref{FLLrho}) as
\begin{eqnarray}
\forall \rho \in {\cal R}^{\B}, \; F[\rho] = F^{\B}[\rho] + \bar{E}^{\B}[\rho],
\label{FLLrhodecomp}
\end{eqnarray}
where $\bar{E}^{\B}[\rho]$ is the complementary basis-set correction density functional
\begin{eqnarray}
\bar{E}^{\B}[\rho] = \braket{\Psi[\rho]}{(T + W_\ee)\Psi[\rho]} - \braket{\Psi^{\B}[\rho]}{(T + W_\ee)\Psi^{\B}[\rho]},
\nonumber\\
\label{}
\end{eqnarray}
and $\Psi[\rho]$ and $\Psi^{\B}[\rho]$ are minimizing wave functions in Eqs.~(\ref{FLLrho}) and~(\ref{FLLBrho}), respectively. Clearly, since ${\cal W}^\B_\rho \subset {\cal W}_\rho$, we have $\forall \rho \in {\cal R}^\B, F^\B[\rho] \geq F[\rho]$, and thus $\bar{E}^{\B}[\rho]\leq 0$. Since the decomposition in Eq.~(\ref{FLLrhodecomp}) is defined only for $\rho \in {\cal R}^{\B}$, we cannot recover the exact ground-state energy $E_0[v]$. Instead, we can obtain the following approximate energy obtained by restricting the minimization in Eq.~(\ref{E0vrhoLLRN}) to densities $\rho$ in ${\cal R}^{\B}$
\begin{eqnarray}
E_{0}^{\B}[v] &=& \min_{\rho \in {\cal R}^{\B}} \left( F[\rho] +  (v,\rho) \right)
\nonumber\\
       &=& \min_{\rho \in {\cal R}^{\B}} \left( \min_{\Psi \in {\cal W}^{\B}_\rho} \braket{\Psi}{(T + W_\ee)\Psi} + \bar{E}^{\B}[\rho] + (v,\rho) \right)
\nonumber\\
&=& \min_{\Psi \in {\cal W}^{\B}} \left( \braket{\Psi}{(T + W_\ee + V)\Psi} + \bar{E}^{\B}[\rho_\Psi] \right),
\label{E0BvrhodecompLL}
\end{eqnarray}
or, designating by $\Psi_0^\B \in {\cal W}^{\B}$ a minimizing wave function in Eq.~(\ref{E0BvrhodecompLL}),
\begin{eqnarray}
E_{0}^{\B}[v] &=& \braket{\Psi_0^\B}{(T + W_\ee + V)\Psi_0^\B} + \bar{E}^{\B}[\rho_{\Psi_0^\B}].
\label{E0Bmin}
\end{eqnarray}
It is easy to see that $E_0[v] \leq E_{0}^{\B}[v] \leq E_{\FCI}^{\B}[v]$. For a given basis set $\B$, the functional $\bar{E}^{\B}[\rho]$ provides a (self-consistent) basis-set correction to the FCI energy so that $E_{0}^{\B}[v]$ is a better approximation to $E_0[v]$ than $E_{\FCI}^{\B}[v]$ is. Moreover, as the basis set is increased toward completeness, $E_{0}^{\B}[v]$ should converge much faster to $E_0[v]$ than $E_{\FCI}^{\B}[v]$ does, since, roughly speaking, densities $\rho$ typically converges faster than wave functions $\Psi$ with respect to the basis set. 

For simplicity, instead of performing the minimization in Eq.~(\ref{E0BvrhodecompLL}), one may use a non-self-consistent approximation consisting in using the FCI ground-state wave function $\Psi_\FCI^\B$ in place of $\Psi_0^\B$, giving what we will call a ``FCI+DFT'' energy
\begin{eqnarray}
E_\text{FCI+DFT}^{\B}[v] &=& \braket{\Psi_\FCI^\B}{(T + W_\ee + V)\Psi_\FCI^\B} + \bar{E}^{\B}[\rho_{\Psi_\FCI^\B}], \;\;\;
\label{EFCI+DFTB}
\end{eqnarray}
\vspace{0.1cm}
which is an upper bound of $E_{0}^{\B}[v]$, i.e. $E_\text{FCI+DFT}^{\B}[v] \geq E_{0}^{\B}[v]$. Again, as the basis set is increased toward completeness, $E_\text{FCI+DFT}^{\B}[v]$ should converge much faster to $E_0[v]$ than $E_{\FCI}^{\B}[v]$ does.

One inconvenient of this basis-set correction scheme is that, for a given finite basis set $\B$, it does not give the exact ground-state energy, even in principle if we knew the exact complementary basis-set correction functional $\bar{E}^{\B}[\rho]$. This is due to the fact that $F^{\B}[\rho]$ is defined only on the restricted set of densities ${\cal R}^{\B}$. Another related inconvenient is that since $\bar{E}^{\B}[\rho]$ is defined only on this restricted set of densities, it is not clear how to define the LDA for it. Defining the LDA would indeed require to consider uniform densities, but uniform densities are not in ${\cal R}^{\B}$. Even though uniform densities are not in ${\cal R}$ either, they can be approached with densities from ${\cal R}$~\cite{LewLieSei-JEP-18,LewLieSei-PAA-20}, so it would be preferable to have a complementary basis-set correction functional defined on the entire set ${\cal R}$. This would also permit to connect in principle the basis-set correction scheme to the exact ground-state energy. This is what is achieved by the second variant of basis-set correction.

\subsection{Second variant of basis-set correction}
\label{sec:secondvariant}

For the second variant of basis-set correction, we work on the full Hilbert space ${\cal H}$ (not restricted to the basis set $\B$), and define the following Hamiltonian
\begin{eqnarray}
H^{\w\B}[v] = T + W_\ee^\B + V,
\label{HwBv}
\end{eqnarray}
where the kinetic-energy operator $T$ and the external potential operator $V$ are still defined as before, but the electron-electron interaction operator is now projected in the basis set $\B$
\begin{eqnarray}
W_\ee^\B = P^\B W_\ee P^\B,
\label{}
\end{eqnarray}
where $P^\B$ is the orthogonal projector onto the basis-set-restricted Hilbert space ${\cal H}^{\B}$. The notation ``\w\B'' is to remind us that only $W_\ee$ is projected. Of course, $W_\ee^\B$ is a complicated non-local two-electron operator. Using an orthonormal orbital basis $\{ \phi_i \}_{i=1,...,M}$ spanning the same space as $\B$, its integral kernel can be written as
\begin{widetext}
\begin{eqnarray}
W_\ee^\B(x_1,x_2;x_1',x_2') = \sum_{i=1}^M \sum_{j=1}^M \sum_{k=1}^M \sum_{l=1}^M  \phi_i(x_1) \phi_j(x_2)\; \braket{\phi_{i}\phi_{j}}{W_\ee \phi_{k}\phi_{l}}\; \phi_k^*(x_1') \phi_l^*(x_2'),
\label{WeeBkernel}
\end{eqnarray}
\end{widetext}
where $\braket{\phi_{i}\phi_{j}}{W_\ee \phi_{k}\phi_{l}}=\int_{\mathbb{R}} \phi_{i}^*(x)\phi_{j}^*(x)\phi_{k}(x) \phi_{l}(x)\d x$ are the two-electron integrals in the orbital basis $\{ \phi_i \}$. For $v \in {\cal V}$, the associated ground-state energy is
\begin{eqnarray}
E_{0}^{\w\B}[v] = \inf_{\Psi \in {\cal W}} \braket{\Psi}{H^{\w\B}[v]\Psi}.
\label{E0FCIW}
\end{eqnarray}
Clearly, if we were to restrict the minimization in Eq.~(\ref{E0FCIW}) to the set ${\cal W}^{\B}$, $E_{0}^{\w\B}[v]$ would reduce to $E_{\text{FCI}}^{\B}[v]$ [Eq.~(\ref{EFCIBv})]. Therefore, we have $E_{0}^{\w\B}[v] \leq E_{\text{FCI}}^{\B}[v]$. Moreover, because $W_\ee$ is a positive operator, one would expect that projecting it in a finite basis should typically decrease the ground-state energy, i.e. $E_{0}^{\w\B}[v] \leq E_0[v]$, but this may not be generally true.

We then define the corresponding Levy-Lieb functional for all densities $\rho \in {\cal R}$ as
\begin{eqnarray}
\forall \rho \in {\cal R}, \; F^{\w\B}[\rho] = \min_{\Psi \in {\cal W}_\rho} \braket{\Psi}{(T + W_\ee^\B)\Psi}.
\label{FLLwB}
\end{eqnarray}
For the same reasons as before, comparison with Eq.~(\ref{FLLBrho}) shows that, for $\rho \in {\cal R}^{\B}$, $F^{\w\B}[\rho] \leq F^{\B}[\rho]$. The ground-state energy $E_{0}^{\w\B}[v]$ can be written as
\begin{eqnarray}
E_{0}^{\w\B}[v] = \inf_{\rho \in {\cal R}} \left( F^{\w\B}[\rho] + (v,\rho) \right).
\label{}
\end{eqnarray}

We now decompose the exact Levy-Lieb density functional $F[\rho]$ as
\begin{eqnarray}
\forall \rho \in {\cal R}, \; F[\rho] = F^{\w\B}[\rho] + \bar{E}_\Hxc^{\w\B}[\rho],
\label{FdecompwB}
\end{eqnarray}
which defines the complementary Hartree-exchange-correlation (Hxc) basis-set correction functional $\bar{E}_\Hxc^{\w\B}[\rho]$. Analogously to what is done in multideterminant range-separated DFT~\cite{TouGorSav-TCA-05,GorSav-IJQC-09a,StoTeaTouHelFro-JCP-13,RebTouTeaHelSav-MP-15,RebTeaHelSavTou-MP-18,FerGinTou-JCP-19}, the functional $\bar{E}_\Hxc^{\w\B}[\rho]$ can be decomposed as
\begin{eqnarray}
\bar{E}_\Hxc^{\w\B}[\rho] = \bar{E}_{\Hx,\md}^{\w\B}[\rho] + \bar{E}_{\c,\md}^{\w\B}[\rho],
\label{FwBdecomp}
\end{eqnarray}
where $\bar{E}_{\Hx,\md}^{\w\B}[\rho]$ is the Hartree-exchange (Hx) contribution defined as the expectation value of the complementary interaction $\bar{W}_\ee^\B = W_\ee - W_\ee^\B$ over the minimizing multideterminant (md) wave function $\Psi^{\w\B}[\rho]$ (that we will assume to be unique up to a global phase factor) in Eq.~(\ref{FLLwB})
\begin{eqnarray}
\bar{E}_{\Hx,\md}^{\w\B}[\rho] = \braket{\Psi^{\w\B}[\rho]}{\bar{W}_\ee^\B \Psi^{\w\B}[\rho]},
\label{EHxmdwB}
\end{eqnarray}
and $\bar{E}_{\c,\md}^{\w\B}[\rho]$ is the remaining correlation (c) contribution
\begin{eqnarray}
\bar{E}_{\c,\md}^{\w\B}[\rho] &=& \braket{\Psi[\rho]}{(T + W_\ee) \Psi[\rho]} 
\nonumber\\
&&- \braket{\Psi^{\w\B}[\rho]}{(T + W_\ee) \Psi^{\w\B}[\rho]}.
\label{EcmdwB}
\end{eqnarray}
Clearly, since $\Psi[\rho]$ minimizes $\braket{\Psi}{(T + W_\ee) \Psi}$, we have $\bar{E}_{\c,\md}^{\w\B}[\rho] \leq 0$. Since the decomposition in Eq.~(\ref{FdecompwB}) is defined for all densities $\rho \in {\cal R}$, we can obtain the exact ground-state energy as
\begin{eqnarray}
E_0[v] &=& \inf_{\rho \in {\cal R}} \left( F^{\w\B}[\rho] + \bar{E}_\Hxc^{\w\B}[\rho] + (v,\rho) \right)
\nonumber\\
       &=& \inf_{\rho \in {\cal R}} \left( \min_{\Psi \in {\cal W}_\rho} \braket{\Psi}{(T + W_\ee^\B)\Psi} + \bar{E}_\Hxc^{\w\B}[\rho] + (v,\rho) \right)
\nonumber\\
       &=& \inf_{\Psi \in {\cal W}} \left( \braket{\Psi}{(T + W_\ee^\B + V)\Psi} + \bar{E}_\Hxc^{\w\B}[\rho_\Psi] \right).
\label{E0vfrom2}
\end{eqnarray}
For potentials $v$ for which there exists a minimizing wave function $\Psi_0^{\w\B} \in {\cal W}$ in Eq.~(\ref{E0vfrom2}), this wave function yields a exact ground-state density $\rho_0$, i.e. $\rho_{\Psi_0^{\w\B}} = \rho_0$. It is in fact the minimizing wave function in Eq.~(\ref{FLLwB}) giving the density $\rho_0$, i.e. $\Psi_0^{\w\B}=\Psi^{\w\B}[\rho_0]$. Using this fact and combining Eqs.~(\ref{E0vfrom2}),~(\ref{FwBdecomp}), and~(\ref{EHxmdwB}), we can express the exact ground-state energy as
\begin{eqnarray}
E_{0}[v] &=& \braket{\Psi_0^{\w\B}}{(T + W_\ee + V)\Psi_0^{\w\B}} + \bar{E}^{\w\B}_{\c,\md}[\rho_{\Psi_0^{\w\B}}].
\label{E0vPsi0wB}
\end{eqnarray}
Thus, this second variant of basis-set correction leads to an energy expression similar to Eq.~(\ref{E0Bmin}) obtained for the first variant of basis-set correction, with the functional $\bar{E}^{\w\B}_{\c,\md}[\rho]$ replacing the functional $\bar{E}^{\B}[\rho]$ and the wave function $\Psi_0^{\w\B}$ replacing the wave function $\Psi_0^{\B}$. One advantage of this second variant of basis-set correction is that it gives the exact ground-state energy $E_{0}[v]$. The price to pay is that the minimization in Eq.~(\ref{E0vfrom2}) is more complicated than in Eq.~(\ref{E0BvrhodecompLL}) since it is over general wave functions $\Psi \in {\cal W}$ and not simply wave functions restricted to ${\cal W}^\B$. Moreover, the minimization involves not only the functional $\bar{E}^{\w\B}_{\c,\md}[\rho]$ but also the functional $\bar{E}_{\Hx,\md}^{\w\B}[\rho]$ in Eq.~(\ref{EHxmdwB}) which did not appear in the first variant of basis-set correction.

Similarly to the first variant of basis-set correction, we can define a non-self-consistent approximation consisting in using the FCI ground-state wave function $\Psi_\FCI^\B$ in place of $\Psi_0^{\w\B}$ in Eq.~(\ref{E0vPsi0wB}), giving an alternative ``FCI+DFT'' energy
\begin{eqnarray}
E_\text{FCI+DFT}^{\w\B}[v] &=& \braket{\Psi_\FCI^{\B}}{(T + W_\ee + V)\Psi_\FCI^{\B}} + \bar{E}^{\w\B}_{\c,\md}[\rho_{\Psi_\FCI^{\B}}], \;\;\;\;\;
\label{EFCI+DFTwB}
\end{eqnarray}
which is quite similar but not equivalent to Eq.~(\ref{EFCI+DFTB}) since the complementary basis-set correction functional is different. Like for the first variant of basis-set correction, when the basis set is increased toward completeness, $E_\text{FCI+DFT}^{\w\B}[v]$ should converge much faster to $E_0[v]$ than $E_{\FCI}^{\B}[v]$ does.

Finally, we define the Lieb density functional for this second variant of basis-set correction
\begin{eqnarray}
\forall \rho \in {\cal R},\; F^{\w\B}_\text{L}[\rho] = \sup_{v \in {\cal V}} \left( E_{0}^{\w\B}[v] - (v,\rho) \right).
\label{FwBL}
\end{eqnarray}
Like in standard DFT, by the theory of Legendre–Fenchel transformations, this Lieb functional $F^{\w\B}_\text{L}$ must be the lower semi-continuous convex envelope of the Levy-Lieb functional $F^{\w\B}$, i.e.
\begin{eqnarray}
F_\text{L}^{\w\B} = \text{lscv}(F^{\w\B}) \leq F^{\w\B}.
\label{FDMLLLwB}
\end{eqnarray}
One could also write down this Lieb functional as a constrained-search over ensemble density matrices, and again we should have $F^{\w\B}[\rho] = F_\text{L}^{\w\B}[\rho]$ for densities $\rho$ which are densities of a non-degenerate ground state of the Hamiltonian $H^{\w\B}[v]$ for some potential $v$. As already mentioned, in the present case of two spin-singlet electrons, the ground state is always non-degenerate and thus the Levy-Lieb and Lieb functionals are identical, i.e. $F^{\w\B}=F_\text{L}^{\w\B}$. As we will see in Section~\ref{sec:fUEGwB}, the definition in Eq.~(\ref{FwBL}) is useful to calculate the functional in practice since it involves a unconstrained maximization over potentials $v$ whereas the definition in Eq.~(\ref{FLLwB}) involves a potentially more complicated constrained minimization over wave functions yielding a fixed density $\rho$.

In summary, the advantage of the second variant of basis-set correction over the first variant is that it is connected with the exact ground-state energy [via Eq.~(\ref{E0vfrom2}) or ~(\ref{E0vPsi0wB})] and that it involves a complementary basis-set correction functional which is defined for all densities in ${\cal R}$. In the next section, we exploit this in order to construct a LDA for the functional $\bar{E}^{\w\B}_{\c,\md}[\rho]$.

\section{Local-density approximation from finite uniform-electron gas}
\label{sec:LDA}

In standard DFT, the LDA is based on the infinite UEG. Essentially, for the 1D case, calculating the energy per particle of the infinite UEG amounts to plugging an uniform density $\rho_\text{unif}: x \mapsto \rho_0 \in (0,+\infty)$ in the density functional $F[\rho]$ and taking the thermodynamic limit, i.e. $\lim_{N\to\infty} F[\rho_\text{unif}]/N$. One difficulty is that a non-zero uniform density function $\rho_\text{unif}$ defined on the entire real line $\mathbb{R}$ is obviously not $N$-representable. For the 3D case, the infinite UEG was rigorously mathematically defined in Refs.~\onlinecite{LewLieSei-JEP-18,LewLieSei-PAA-20} by first convoluting the uniform density with a function of compact support (so that the convoluted density is $N$-representable) and then taking the thermodynamic limit $N\to\infty$ (after removing the Hartree energy which is divergent for the Coulombic 3D case). Here, in the spirit of Ref.~\onlinecite{GilLoo-TCA-12}, we will instead consider a \textit{finite} UEG, i.e., for a finite electron number $N$.

\subsection{Finite uniform-electron gas for the complete-basis-set case}
\label{sec:fUEGcbs}

To define a 1D finite UEG, we generalize the standard DFT of Section~\ref{sec:standardDFT} from the real line $\mathbb{R}$ to a finite interval $\Omega_a=(-a/2,a/2)$ of length $a$. Hence, the one-electron Hilbert space is $\calh_a = L^2 (\Omega_a, \mathbb{C})$ and the two-electron Hilbert space is ${\cal H}_a = \calh_a \otimes \calh_a$. For external local potentials $v \in {\cal V}_a$ (where the space ${\cal V}_a$ will be specified below) and $N=2$ electrons, we define the ground-state energy of the Hamiltonian $H[v]=T+W_\ee+V$ [Eq.~(\ref{Hv})] restricted to the Hilbert space ${\cal H}_a$ as
\begin{eqnarray}
E_{0,a}[v] = \inf_{\Psi \in {\cal W}_a} \braket{\Psi}{H[v]\Psi}_a,
\label{}
\end{eqnarray}
with the set of admissible wave functions
\begin{equation}
{\cal W}_a = \Big\{ \Psi \in {\cal H}_a\;|\; \Psi \in H^1_\text{per}(\Omega_a^2, \mathbb{C}),\; \braket{\Psi}{\Psi}_a=1 \Big\},
\label{}
\end{equation}
where $\braket{\Psi_1}{\Psi_2}_a=\int_{\Omega_a^2} \Psi_1^*(x_1,x_2) \Psi_2(x_1,x_2)\d x_1 \d x_2$ is the inner product on ${\cal H}_a$ and $H^1_\text{per}$ designates the set of functions in $H^1$ with periodic boundary conditions on the domain (see, e.g., Ref.~\onlinecite{Lew-INC-21}), which can be defined as $H^1_\text{per}(\Omega_a^2, \mathbb{C})=\left\{ \Psi_{|\Omega_a^2} \; |\; \Psi \in H^1_\text{loc}(\mathbb{R}^2, \mathbb{C}), \; \Psi\; \text{is}\; a\mathbb{Z}^2\text{-periodic} \right\}$ where $\Psi_{|\Omega_a^2}$ designates the restriction of $\Psi$ to $\Omega_a^2$ and $H^1_\text{loc}(\mathbb{R}^2, \mathbb{C})$ is the local first-order Sobolev space.

The corresponding Levy-Lieb density functional is
\begin{eqnarray}
\forall \rho \in {\cal R}_a, \; F_{a}[\rho] = \min_{\Psi \in {\cal W}_{a,\rho}} \braket{\Psi}{(T + W_\ee)\Psi}_a,
\label{}
\end{eqnarray}
where ${\cal W}_{a,\rho} = \left\{ \Psi \in {\cal W}_a,\; \rho_{\Psi} = \rho \right\}$ and ${\cal R}_a$ is the set of $N$-representable densities on $\Omega_a$
\begin{eqnarray}
{\cal R}_a &=& \left\{ \rho \; | \; \exists \; \Psi \in {\cal W}_a, \rho_\Psi  = \rho \right\} 
\nonumber\\
&=& \left\{\rho \in L^1(\Omega_a) \;|\; \rho \geq 0, \; \int_{\Omega_a} \rho(x) \d x = N, \; \sqrt{\rho} \in H^1_\text{per}(\Omega_a) \right\},
\nonumber\\
\label{}
\end{eqnarray}
and $H^1_\text{per}(\Omega_a)=\{f \in H^1(\Omega_a) \; | \; \lim_{x\to -a/2} f(x)= \lim_{x\to a/2}f(x) \}$. We have ${\cal R}_a \subset {\cal X}_a$ where ${\cal X}_a$ is the Banach space ${\cal X}_a = C_\text{per}(\Omega_a) \cap L^1(\Omega_a)$ with $C_\text{per}(\Omega_a)$ the space of continuous functions on $\Omega_a$ with periodic boundary conditions. Therefore, the space of external potentials to consider is the continuous dual space of ${\cal X}_a$, i.e. ${\cal V}_a= {\cal X}_a' = M_\text{per}(\Omega_a) + L^\infty(\Omega_a)$ where $M_\text{per}(\Omega_a)$ is the space of bounded Radon measures on $\Omega_a$ with periodic boundary conditions. In the limit of an infinite interval ($a \to \infty$), we recover the standard theory on the real line $\mathbb{R}$, i.e. $\lim_{a\to \infty} F_{a}[\rho] = F[\rho]$ where $F[\rho]$ is the standard Levy-Lieb functional defined in Eq.~(\ref{FLLrho}). Similarly, we could generalize the Lieb density functionals in Eq.~(\ref{FLrho}) to the finite interval $\Omega_a$.

We now define a \textit{finite} UEG (fUEG) by considering the uniform density $\rho_\text{unif}: x \mapsto \rho_0=N/a$ on the interval $\Omega_a$ for the fixed electron number $N=2$. Note that $\rho_\text{unif}$ is in fact the unique uniform density belonging to ${\cal R}_a$. The energy per particle of this finite UEG is
\begin{eqnarray}
\varepsilon_\text{fUEG}(\rho_0) = \frac{F_{a}[\rho_\text{unif}]}{N},
\label{epsfUEG}
\end{eqnarray}
and is a function of the only variable $\rho_0$ since $N$ is fixed and $a=N/\rho_0$. 
The value of $F_{a}[\rho_\text{unif}]$ corresponds to the ground-state energy of the two-electron Hamiltonian with zero external potential
\begin{eqnarray}
H_\text{fUEG} = H[0] = T + W_\ee,
\label{HfUEG}
\end{eqnarray}
with periodic boundary conditions on $\Omega_a$, provided that the ground-state density is the uniform density $\rho_\text{unif}$ (i.e., no translational symmetry breaking). We note that in Refs.~\onlinecite{LooGil-PRL-12,LooGil-JCP-13,LooBalGil-JCP-14,Loo-PRA-14} 1D finite UEGs mapped to a ring were introduced using the Coulomb electron-electron interaction. Here, instead we use the Dirac-delta electron-electron interaction and we do not work on a ring.

For a given density $\rho_0$ in the range $[0,10]$ a.u. and for the fixed electron number $N=2$ and interval length $a=N/\rho_0$, we calculate the ground-state energy by performing a FCI calculation using a one-electron plane-wave (pw) orthonormal basis $\{ p_n \}_{n\in \mathbb{Z},\; |n|\leq n_\text{max}^\text{pw}}$ where $p_n(x) = (1/\sqrt{a}) e^{\i k_n x}$ and $k_n = 2\pi n /a$. The one-electron kinetic integrals $\braket{p_{n_1}}{t p_{n_2}}_a = (2\pi^2 n_1^2/a^2) \delta_{n_1,n_2}$ and the two-electron integrals $\braket{p_{n_1}p_{n_2}}{W_\ee p_{n_3}p_{n_4}}_a = (1/a) \delta_{n_1+n_2,n_3+n_4}$ are trivial. We use a plane-wave cutoff $n_\text{max}^\text{pw}=60$ which leads to FCI energies converged to at least 1 mhartree (and in fact generally better than that). As usual, the finite UEG energy per particle can be decomposed as
\begin{eqnarray}
\varepsilon_\text{fUEG}(\rho) = t_{\s,\text{fUEG}}(\rho) + \varepsilon_{\H,\text{fUEG}}(\rho) + \varepsilon_{\x,\text{fUEG}}(\rho) + \varepsilon_{\c,\text{fUEG}}(\rho),
\nonumber\\
\label{epscfUEGdecomp}
\end{eqnarray}
with the non-interacting kinetic energy per particle $t_{\s,\text{fUEG}}(\rho)=0$ (since $N=2$ the only occupied orbital is the constant plane wave $p_0(x)=1/\sqrt{a}$ which has a zero kinetic energy), the Hartree energy per particle $\varepsilon_{\H,\text{fUEG}}(\rho)=\rho/2$, the exchange energy per particle $\varepsilon_{\x,\text{fUEG}}(\rho)=-\rho/4$, and the correlation energy per particle $\varepsilon_{\c,\text{fUEG}}(\rho)$.

\begin{figure}
\includegraphics[scale=0.35,angle=-90]{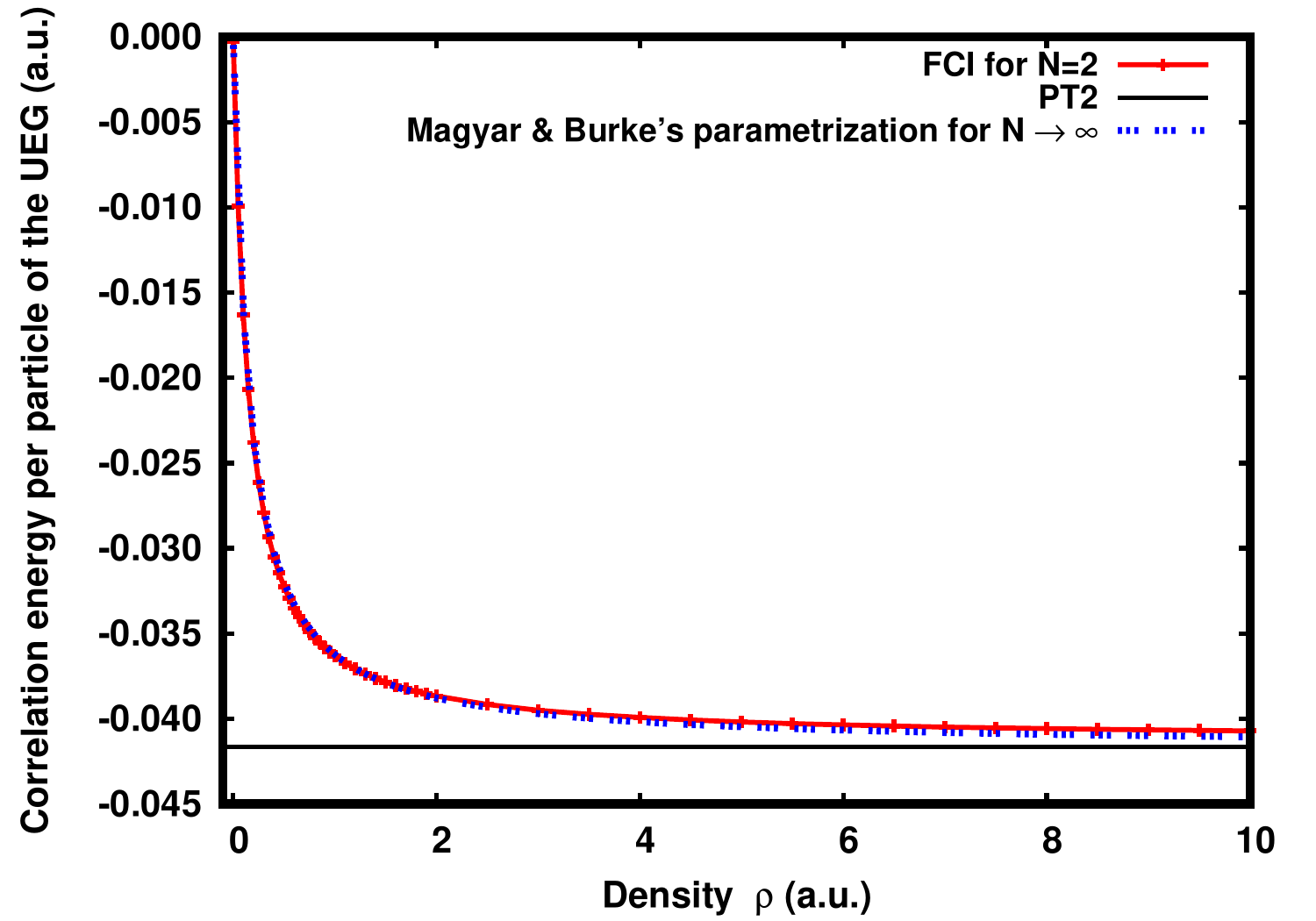}
\caption{FCI correlation energy per particle $\varepsilon_{\c,\text{fUEG}}(\rho)$ [Eqs.~(\ref{epsfUEG}) and~(\ref{epscfUEGdecomp})] of the 1D finite UEG as a function of the density $\rho$ for $N=2$ electrons and a plane-wave cutoff $n_\text{max}^\text{pw}=60$. The exact PT2 correlation energy per particle, which is independent of $N$ and $\rho$, is indicated as a horizontal line. The correlation energy per particle of the infinite UEG ($N\to\infty$) as parametrized by Magyar and Burke~\cite{MagBur-PRA-04} from essentially numerically exact Bethe-ansatz results is also plotted for comparison.}
\label{fig:fUEGepsc}
\end{figure}

The correlation energy per particle $\varepsilon_{\c,\text{fUEG}}(\rho)$ is plotted in Fig.~\ref{fig:fUEGepsc}. In the high-density limit, $\varepsilon_{\c,\text{fUEG}}(\rho)$ tends to the correlation energy per particle in second-order perturbation theory (PT2) with respect to the electron-electron interaction $W_\ee$
\begin{eqnarray}
\lim_{\rho \to\infty} \varepsilon_{\c,\text{fUEG}}(\rho)  = \varepsilon_{\c,\text{fUEG}}^\text{PT2} = -\frac{1}{24},
\label{epscfUEGrhoinf}
\end{eqnarray}
which is a constant independent of $\rho$. It turns out that this constant is the same for $N=2$ and $N\to \infty$~\cite{MagBur-PRA-04}. In the low-density limit, $\varepsilon_{\c,\text{fUEG}}(\rho)$ goes to zero linearly with $\rho$ (see Ref.~\onlinecite{RasSeiGor-PRB-11})
\begin{eqnarray}
\varepsilon_{\c,\text{fUEG}}(\rho)  \isEquivTo{\rho \to 0} -\frac{\rho}{4},
\label{epscfUEGrho0}
\end{eqnarray}
so as to exactly cancel out the Hartree and exchange energies per particle. This is due to the fact that, in this limit, the probability density of finding the electrons at the same point of space is zero and thus the Dirac-delta electron-electron interaction has not effect. This is the 1D version of the strong-interaction limit of DFT~\cite{SeiPerLev-PRA-99,Sei-PRA-99,SeiGorSav-PRA-07,GorSei-PCCP-10}. Equation~(\ref{epscfUEGrho0}) is also true for $N\to \infty$~\cite{MagBur-PRA-04,RasPitCapPro-PRL-09}, and is in fact true independently of $N$~\cite{RasSeiGor-PRB-11}. In Fig.~\ref{fig:fUEGepsc}, we also show the correlation energy per particle of the infinite UEG ($N\to\infty$) as parametrized by Magyar and Burke~\cite{MagBur-PRA-04} from essentially numerically exact Bethe-ansatz results. Not only the correlation energies per particle for $N=2$ and $N\to\infty$ agree well for small and large densities, as they should since they have the same $N$-independent asymptotic behaviors [Eqs.~(\ref{epscfUEGrhoinf}) and (\ref{epscfUEGrho0})], but they also agree very well for intermediate densities (the maximal deviation between the two curves being about 0.4 mhartree), showing that the thermodynamic limit $N\to\infty$ is essentially already reached at $N=2$ for this 1D UEG. Hence, there is no need considering 1D UEGs with $N>2$ electrons. This must be due to the very short-range nature of the Dirac-delta electron-electron interaction. For the 1D UEG with the Coulomb interaction, the correlation energy per particle depends much more strongly on the electron number~\cite{LooGil-JCP-13}.

\subsection{Finite uniform-electron gas for the incomplete-basis-set case}
\label{sec:fUEGwB}

We now generalize the second variant of basis-set correction of Section~\ref{sec:secondvariant} from the real line $\mathbb{R}$ to a finite interval $\Omega_a=(-a/2,a/2)$ of length $a$. For $v \in {\cal V}_a$ and $N=2$ electrons, we define the ground-state energy of the restriction to the Hilbert space ${\cal H}_a$ of the Hamiltonian $H^{\w\B}[v]=T+W_\ee^\B +V$ [Eq.~(\ref{HwBv})], featuring the electron-electron interaction projected in the basis set $\B$ used for the 1D He-like atom, as
\begin{eqnarray}
E_{0,a}^{\w\B}[v] = \inf_{\Psi \in {\cal W}_a} \braket{\Psi}{H^{\w\B}[v]\Psi}_a.
\label{}
\end{eqnarray}
The corresponding Levy-Lieb density functional is
\begin{eqnarray}
\forall \rho \in {\cal R}_a, \; F_\text{a}^{\w\B}[\rho] = \min_{\Psi \in {\cal W}_{a,\rho}} \braket{\Psi}{(T + W_\ee^\B)\Psi}_a,
\label{}
\end{eqnarray}
and the corresponding Lieb density functional is
\begin{eqnarray}
\forall \rho \in {\cal R}_a, \; F_\text{L,a}^{\w\B}[\rho] = \sup_{v \in {\cal V}_a} \left( E_{0,a}^{\w\B}[v] - (v,\rho)_a \right),
\label{FLawB}
\end{eqnarray}
where $(v,\rho)_a=\int_{\Omega_a} v(x) \rho(x) \d x$. Again, in the limit of an infinite interval ($a \to \infty$), we recover the theory of Section~\ref{sec:secondvariant}.

For a given basis set $\B$, we now define an associated finite UEG by considering the uniform density $\rho_\text{unif}: x \mapsto \rho_0=N/a$ on the interval $\Omega_a$ for the fixed electron number $N=2$. The kinetic + electron-electron energy per particle of this $\B$-dependent finite UEG is
\begin{eqnarray}
f_\text{fUEG}^{\w\B}(\rho_0) = \frac{F_{a}^{\w\B}[\rho_\text{unif}]}{N},
\label{}
\end{eqnarray}
where
\begin{eqnarray}
F_{a}^{\w\B}[\rho_\text{unif}] = \braket{\Psi^{\w\B}[\rho_\text{unif}]}{(T + W_\ee^{\B})\Psi^{\w\B}[\rho_\text{unif}]}_a,
\label{}
\end{eqnarray}
and $\Psi^{\w\B}[\rho_\text{unif}]$ is the ground-state wave function (assumed to be unique up to a global phase factor) of the two-electron Hamiltonian
\begin{eqnarray}
H_\text{fUEG}^{\w\B} = T + W_\ee^{\B} + V^{\w\B},
\label{HfUEGwB}
\end{eqnarray}
with periodic boundary conditions on $\Omega_a$ and with $V^{\w\B} = \sum_{i=1}^N v^{\w\B}(x_i)$ where $v^{\w\B}(x)$ is the local potential (that we assume to exist and which is defined up to an additive constant) which enforces the constraint that the ground-state wave function $\Psi^{\w\B}[\rho_\text{unif}]$ yields indeed the uniform density $\rho_\text{unif}$. Since the projected electron-electron interaction $W_\ee^{\B}$ breaks translation invariance, the addition of the potential $v^{\w\B}$ is necessary to restore a uniform density. This is in contrast with the UEG for the complete-basis-set case for which no external potential was necessary to obtain a uniform density [Eq.~(\ref{HfUEG})]. To conveniently obtain the potential $v^{\w\B}$, we use the fact that, since the two-electron finite UEG has a non-degenerate ground state, the Levy-Lieb functional $F_{a}^{\w\B}$ and the Lieb functional $F_\text{L,a}^{\w\B}$ are identical.
The potential $v^{\w\B}$ then just corresponds to the maximizing potential in Eq.~(\ref{FLawB}) for $\rho=\rho_\text{unif}$ (see Refs.~\onlinecite{ColSav-JCP-99,WuYan-JCP-03,TeaCorHel-JCP-09})
\begin{eqnarray}
v^{\w\B} = \argmax{v \in {\cal V}_a} \left( E_{0,a}^{\w\B}[v] - (v,\rho_\text{unif})_a \right).
\label{vwBargmax}
\end{eqnarray}

For a given basis set $\B$, for a given density $\rho_0$ in the range $[0,10]$ a.u., and for the fixed electron number $N=2$ and interval length $a=N/\rho_0$, we calculate the energy $E_{0,a}^{\w\B}[v]$  by performing a FCI calculation using a plane-wave orthonormal basis $\{ p_n \}_{n\in \mathbb{Z},\; |n|\leq n_\text{max}^\text{pw}}$. The one-electron kinetic integrals are still $\braket{p_{n_1}}{t p_{n_2}}_a = (2\pi^2 n_1^2/a^2) \delta_{n_1,n_2}$. The integrals of the electron-electron interaction projected in the basis set $\B$ [see Eq.~(\ref{WeeBkernel})] can be calculated as
\begin{widetext}
\begin{eqnarray}
\braket{p_{n_1}p_{n_2}}{W_\ee^\B p_{n_3}p_{n_4}}_a = \sum_{i=1}^M \sum_{j=1}^M \sum_{k=1}^M \sum_{l=1}^M  S_{i,n_1}^* S_{j,n_2}^* \; \braket{\phi_{i}\phi_{j}}{W_\ee \phi_{k}\phi_{l}} \; S_{k,n_3} S_{l,n_4},
\label{}
\end{eqnarray}
\end{widetext}
where $\braket{\phi_{i}\phi_{j}}{W_\ee \phi_{k}\phi_{l}}$ are the two-electron integrals in terms of the HF orbitals $\{ \phi_i \}_{i=1,...,M}$ of the 1D He-like atom expanded in the basis set $\B$ (see Section~\ref{sec:FCIB}) and $S_{i,n} = \braket{\phi_i}{p_n}_a = \int_{\Omega_a} \phi_i^*(x) p_n(x) \d x$ are the overlap integrals between the HF orbitals and the plane-wave basis functions. The potential $v$ to optimize in Eq.~(\ref{vwBargmax}) is also expanded on the same plane-wave basis set
\begin{eqnarray}
v(x) = \sum_{n\in\mathbb{Z},|n|\leq n_\text{max}^\text{pw}} c_{n} p_n(x),
\label{vxexpand}
\end{eqnarray}
with coefficients $c_n \in \mathbb{R}$ and we impose $c_{-n} = c_{n}$ in order to have a real-valued and parity-even potential. The one-electron potential integrals, needed to calculate $E_{0,a}^{\w\B}[v]$, are
\begin{eqnarray}
\braket{p_{n_1}}{v p_{n_2}}_a &=& \int_{\Omega_a} p_{n_1}^*(x) v(x) p_{n_2}(x) \d x = \frac{c_{n_1-n_2}}{\sqrt{a}},
\label{}
\end{eqnarray}
and the second term in Eq.~(\ref{vwBargmax}) is simply $(v,\rho_\text{unif})_a = \sqrt{a} c_0 \rho_0$. Finally, for the optimization of the potential, it is useful to have the derivative of $F[v] = E_{0,a}^{\w\B}[v] - (v,\rho_\text{unif})_a$ with respect to the coefficient $c_n$. Using the Hellmann-Feynman theorem, we find
\begin{eqnarray}
\frac{\partial F[v]}{\partial c_n} &=& (p_n,\rho_{\Psi_v})_a - (p_n,\rho_\text{unif})_a
\nonumber\\
&=& \frac{1}{\sqrt{a}} \sum_{n_1\in\mathbb{Z},|n_1|\leq n_\text{max}^\text{pw}} \gamma_{n_1,n_1+n} - \sqrt{a} \rho_0 \delta_{n,0},
\label{}
\end{eqnarray}
where we have used $\rho_{\Psi_v}(x) = \sum_{n_1\in\mathbb{Z},|n_1|\leq n_\text{max}^\text{pw}} \sum_{n_2\in\mathbb{Z},|n_2|\leq n_\text{max}^\text{pw}} \gamma_{n_1,n_2} p_{n_1}(x) p_{n_2}^*(x)$ with the one-particle reduced density matrix $\gamma$ of the ground-state wave function $\Psi_v$ of $H^{\w\B}[v]$. In practice, we use a plane-wave cutoff of $n_\text{max}^\text{pw}=30$. We can use a smaller cutoff than the cutoff used for the complete-basis-set UEG in Section~\ref{sec:fUEGcbs} since the FCI energy $E_{0,a}^{\w\B}[v]$ has a fast convergence with $n_\text{max}^\text{pw}$ due to the presence of the projected electron-electron interaction $W_\ee^\B$. To optimize the coefficients $\{c_n\}$ of the potential, we use the conjugate gradient method~\cite{PreTeuVetFla-BOOK-92}. Since the term $n=0$ in Eq.~(\ref{vxexpand}) is just an arbitrary constant, we keep the coefficient $c_0$ fixed to $0$. With a zero potential $v$, the FCI density $\rho_{\Psi_{v=0}}(x)$ can deviate from the target density $\rho_0$ by about as much as 0.2 a.u. for the basis set $\B$ of smallest size (i.e., $M=1$). With our optimized potentials $v^{\w\B}$, the density $\rho_{\Psi_{v^{\w\B}}}(x)$ deviates from the target density $\rho_0$ to at most about $10^{-4}$ a.u..

As an aside, it might be worthwhile to stress here that in the Hamiltonian in Eq.~(\ref{HfUEGwB}), the kinetic-energy operator $T$ is not projected in the basis set $\B$. We observed that if $T$ is also projected in the basis set $\B$, then the high-lying states of $T$ collapse to the lower part of the spectrum, which inevitably pollutes the nature of the ground state of the finite UEG Hamiltonian and renders numerically impossible to find a potential restoring a uniform density. This is why in the second-variant of basis-set correction in Section~\ref{sec:secondvariant} we have decided to project only the electron-electron interaction $W_\ee$ in the basis set $\B$.

\begin{figure}
\includegraphics[scale=0.35,angle=-90]{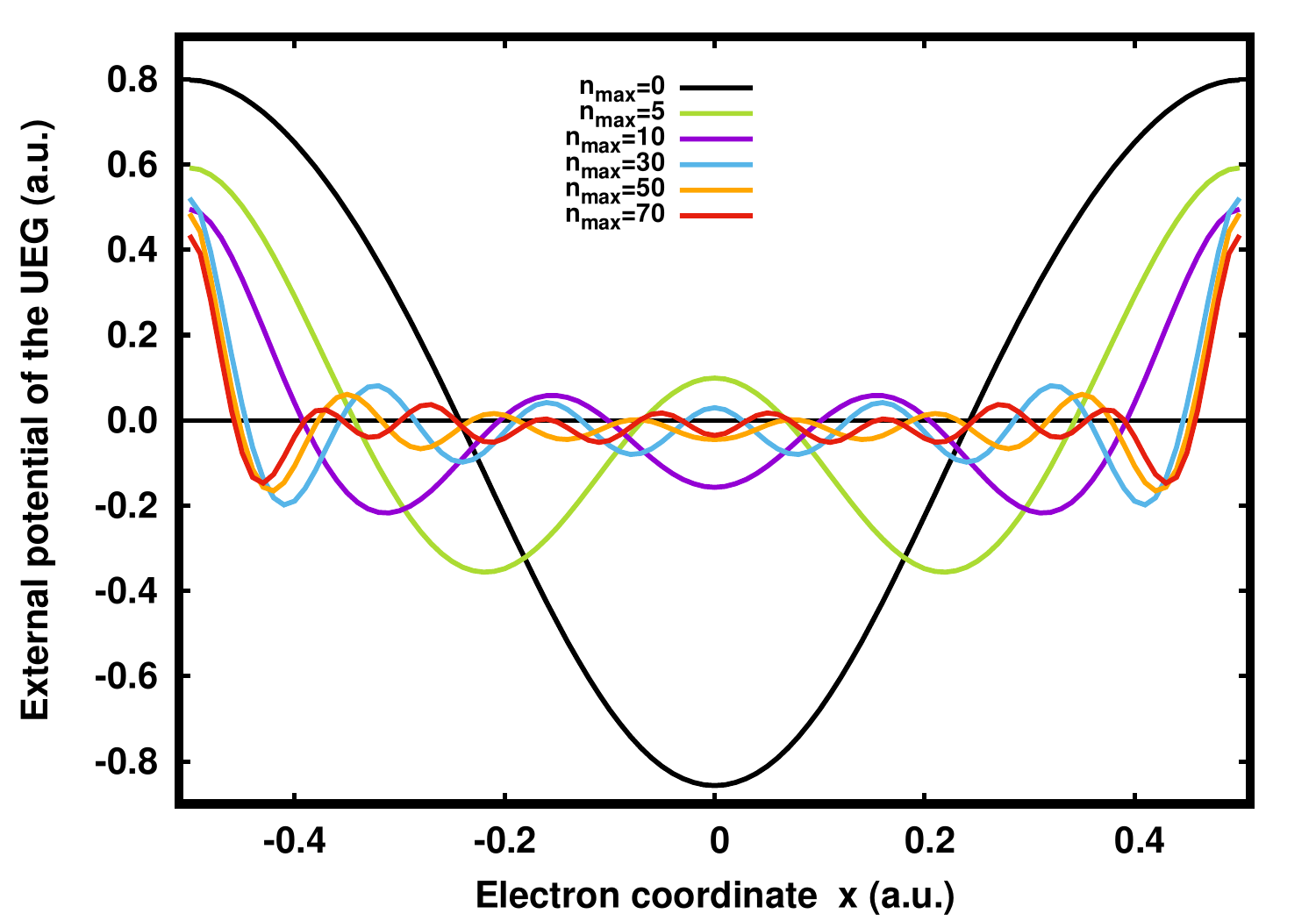}
\caption{External optimized potential $v^{\w\B}(x)$ [Eq.~(\ref{vwBargmax})] keeping a uniform density $\rho_0 = 2$ a.u. for the 1D finite UEG for $N=2$ electrons and for different sizes $n_\text{max}$ of the basis set $\B$ of the 1D He-like atom.}
\label{fig:vwBx}
\end{figure}

The optimized potentials $v^{\w\B}(x)$ obtained from Eq.~(\ref{vwBargmax}) are plotted in Fig.~\ref{fig:vwBx} for the example of the target density $\rho_0=2$ a.u. and for different sizes $n_\text{max}$ of the basis set $\B$ of the 1D He-like atom introduced in Section~\ref{sec:FCIB}. To compensate for the breaking of translation invariance of the projected electron-electron interaction, all potentials show oscillations with maximum amplitude on the edges of the interval. As expected, when $n_\text{max}$ increases, the amplitude of the potential decreases, as it must eventually go to zero in the complete-basis-set limit $n_\text{max} \to \infty$.

\begin{figure}
\includegraphics[scale=0.35,angle=-90]{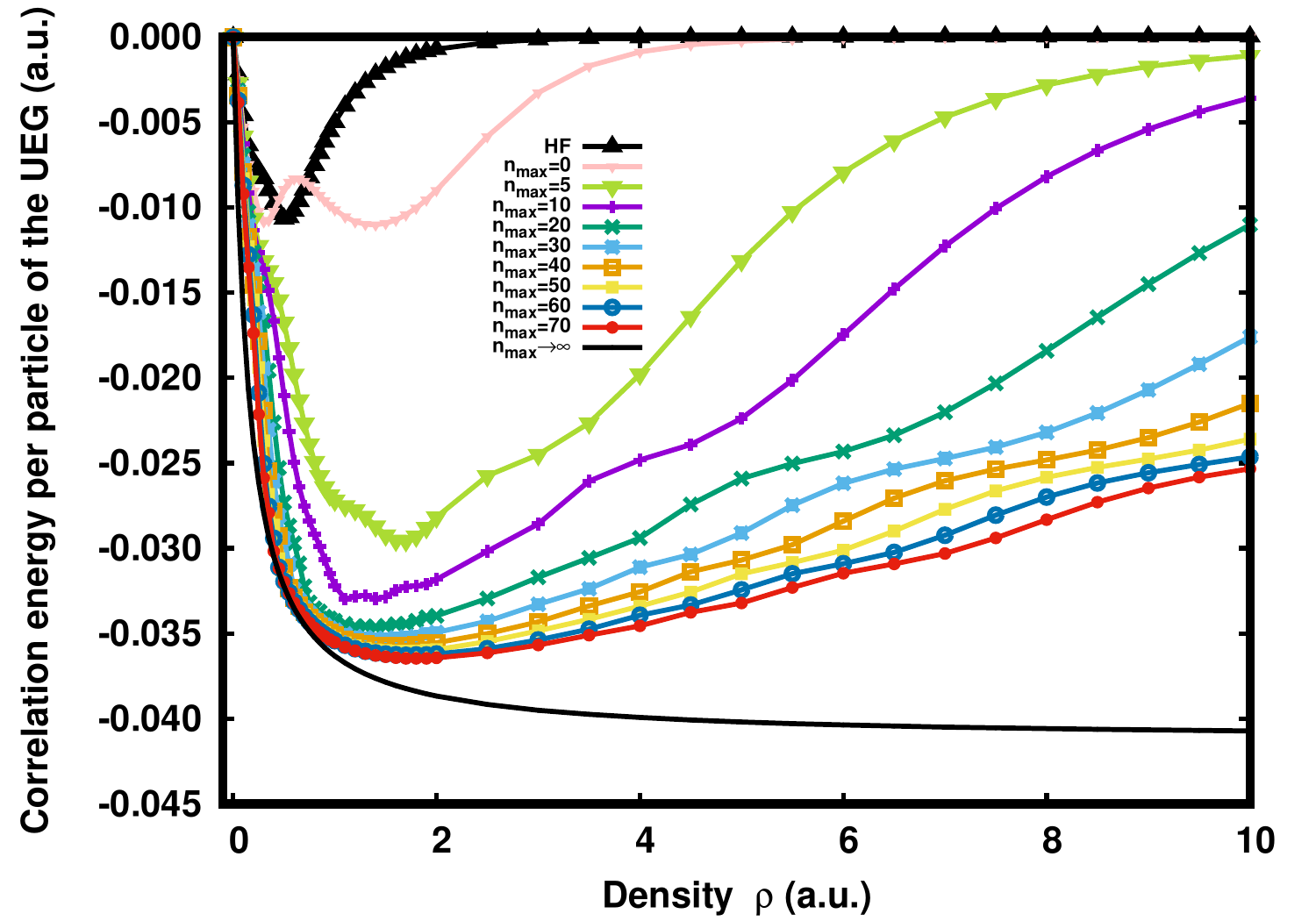}
\caption{FCI correlation energy per particle $\varepsilon_{\c,\text{fUEG}}^{\w\B}(\rho)$ [Eqs.~(\ref{epsfUEGwB}) and~(\ref{epsfUEGwBdecomp})] of the 1D finite UEG as a function of the density $\rho$ for $N=2$ electrons and for different sizes $n_\text{max}$ of the basis set $\B$ of the 1D He-like atom. The curve labelled by ``HF'' corresponds to the limiting case where the basis set $\B$ contains only the exact HF occupied orbital.}
\label{fig:epscfUEGwB}
\end{figure}

Once the FCI ground-state wave function $\Psi^{\w\B}[\rho_\text{unif}]=\Psi_{v^{\w\B}}$ for the optimal potential $v^{\w\B}$ has been obtained, we calculate the following energy per particle using this wave function
\begin{eqnarray}
\varepsilon_\text{fUEG}^{\w\B}(\rho_0) = \frac{\braket{\Psi^{\w\B}[\rho_\text{unif}]}{(T + W_\ee) \Psi^{\w\B}[\rho_\text{unif}]}_a}{N},
\label{epsfUEGwB}
\end{eqnarray}
which we can decompose in the same way as in Eq.~(\ref{epscfUEGdecomp})
\begin{eqnarray}
\varepsilon_\text{fUEG}^{\w\B}(\rho) = t_{\s,\text{fUEG}}(\rho) + \varepsilon_{\H,\text{fUEG}}(\rho) + \varepsilon_{\x,\text{fUEG}}(\rho) + \varepsilon_{\c,\text{fUEG}}^{\w\B}(\rho),
\nonumber\\
\label{epsfUEGwBdecomp}
\end{eqnarray}
with the same kinetic, Hartree, and exchange contributions as in Eq.~(\ref{epscfUEGdecomp}), and a new correlation energy per particle $\varepsilon_{\c,\text{fUEG}}^{\w\B}(\rho)$. This latter quantity is plotted in Fig.~\ref{fig:epscfUEGwB} for different sizes $n_\text{max}$ of the basis set $\B$ of the 1D He-like atom. As expected, when $n_\text{max}$ increases, $\varepsilon_{\c,\text{fUEG}}^{\w\B}(\rho)$ becomes more negative and gets closer to the correlation energy per particle $\varepsilon_{\c,\text{fUEG}}(\rho)$ of the complete-basis-set limit $n_\text{max} \to \infty$. For finite $n_\text{max}$, it can be observed that, in the high-density limit, the correlation energy per particle $\varepsilon_{\c,\text{fUEG}}^{\w\B}(\rho)$ goes to zero, unlike in Eq.~(\ref{epscfUEGrhoinf}). This is due to the fact that, as the density increases, the relevant electron-electron distances contributing to the correlation energy become smaller and the basis set $\B$ is unable to resolve the Dirac-delta electron-electron interaction at a fine enough distance scale.

\begin{figure}
\includegraphics[scale=0.35,angle=-90]{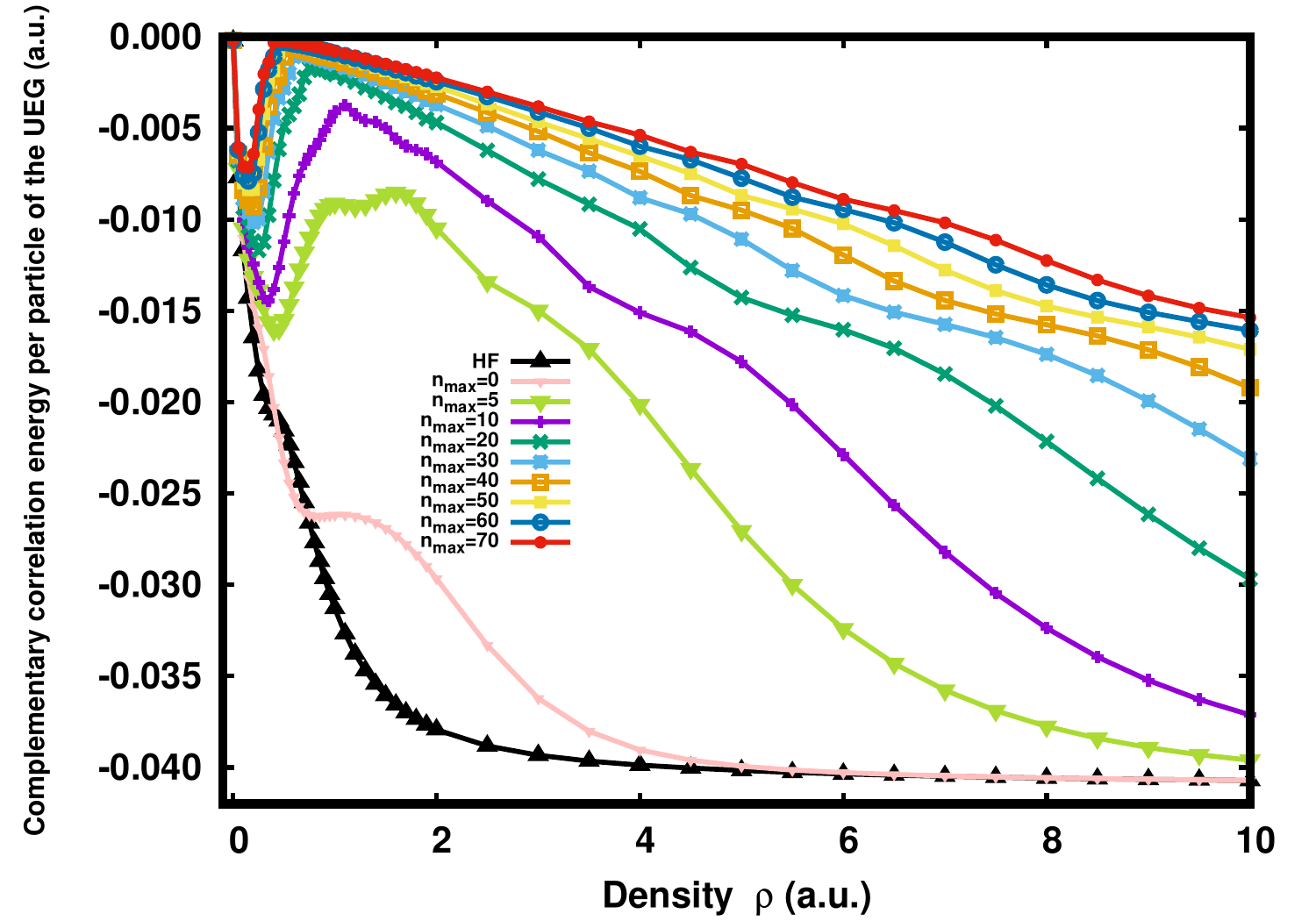}
\caption{FCI complementary correlation energy per particle $\bar{\varepsilon}_{\c,\md,\text{fUEG}}^{\w\B}(\rho)$ [Eq.~(\ref{barepscmdfUEGwB})] of the 1D finite UEG as a function of the density $\rho$ for $N=2$ electrons and for different sizes $n_\text{max}$ of the basis set $\B$ of the 1D He-like atom. The curve labelled by ``HF'' corresponds to the limiting case where the basis set $\B$ contains only the exact HF occupied orbital.}
\label{fig:fUEGbarepscwB}
\end{figure}

Finally, we calculate the complementary multideterminant correlation energy per particle of the finite UEG for the basis set $\B$ [see Eq.~(\ref{EcmdwB})]
\begin{eqnarray}
\bar{\varepsilon}_{\c,\md,\text{fUEG}}^{\w\B}(\rho) = \varepsilon_\text{fUEG}(\rho) - \varepsilon_\text{fUEG}^{\w\B}(\rho),
\label{barepscmdfUEGwB}
\end{eqnarray}
which is plotted in Fig.~\ref{fig:fUEGbarepscwB} for different basis sizes $n_\text{max}$. As $n_\text{max}$ increases, the magnitude of $\bar{\varepsilon}_{\c,\md,\text{fUEG}}^{\w\B}(\rho)$ decreases and must eventually go to zero in the complete-basis-set limit $n_\text{max} \to \infty$. The magnitude is largest for high densities since $\bar{\varepsilon}_{\c,\md,\text{fUEG}}^{\w\B}(\rho)$ must compensate for the inability of the basis set $\B$ to represent the Dirac-delta electron-electron interaction at a small distance scale. Perhaps surprisingly, there is also a local maximum of the magnitude of $\bar{\varepsilon}_{\c,\md,\text{fUEG}}^{\w\B}(\rho)$ at small densities. This is due to the fact that, at small densities, $\bar{\varepsilon}_{\c,\md,\text{fUEG}}^{\w\B}(\rho)$ does not exactly cancel out the Hartree and exchange energies per particle, in contrast to the complete-basis-set case [Eq.~(\ref{epscfUEGrho0})]. Again, this must come from the inability of the basis set $\B$ to represent the Dirac-delta electron-electron interaction sufficiently precisely to give a zero probability density of finding the electrons as the same point of space in the low-density limit. Interestingly, in between the small and the large-density regimes, for $n_\text{max} \geq 5$, the magnitude of $\bar{\varepsilon}_{\c,\md,\text{fUEG}}^{\w\B}(\rho)$ passes through a minimum. In particular, for $n_\text{max}=70$, $\bar{\varepsilon}_{\c,\md,\text{fUEG}}^{\w\B}(\rho)$ is almost zero at around $\rho \approx 0.5$ a.u., which means that the basis set $\B$ accurately captures the effect of the Dirac-delta electron-electron interaction at this density.

\subsection{Finite local-density approximation}

\begin{figure}
\includegraphics[scale=0.35,angle=-90]{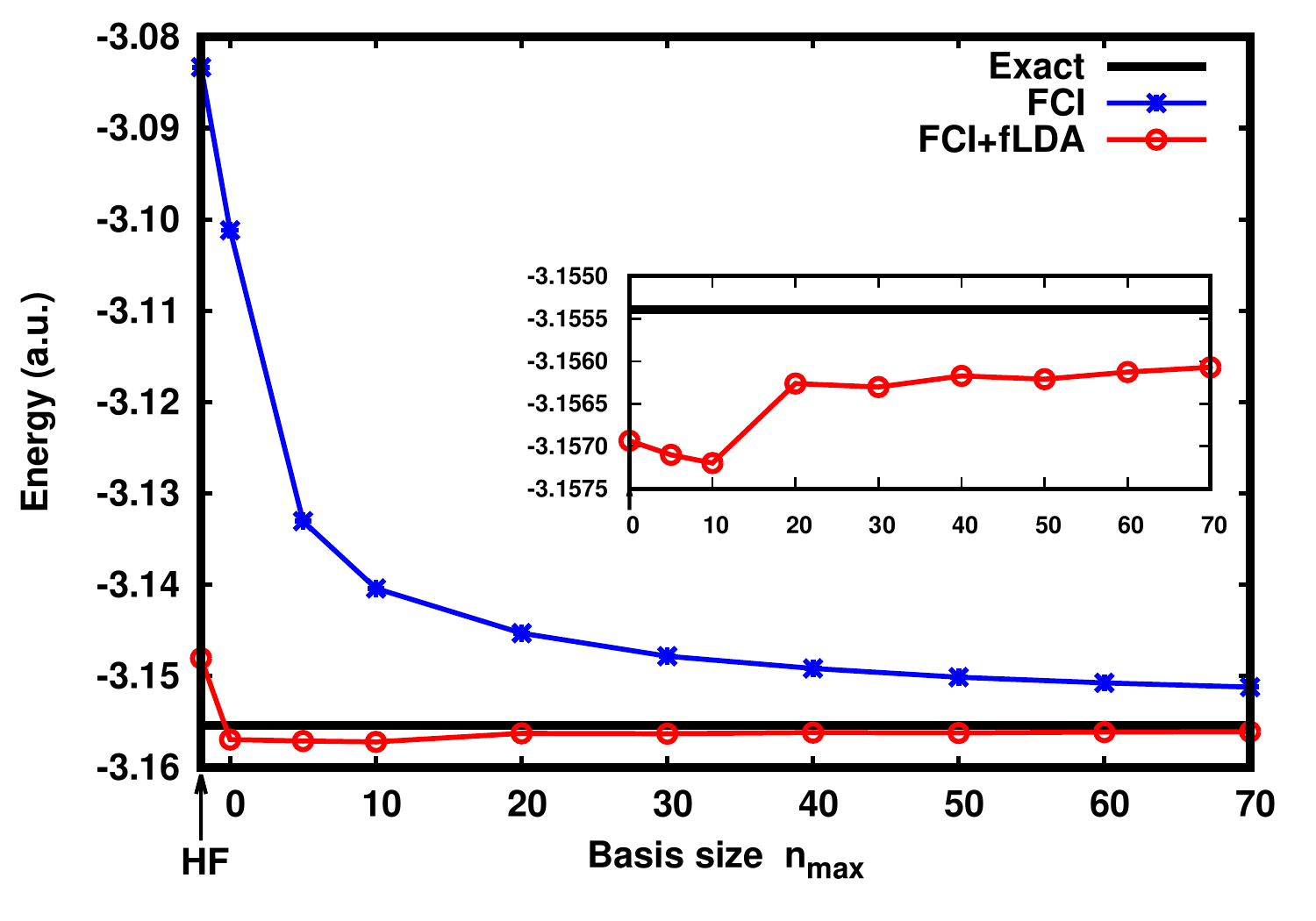}
\caption{FCI ground-state energy $E_{\FCI}^\B$ [Eq.~(\ref{E0BFCI})] and FCI+fLDA ground-state energy $E_{\text{FCI+fLDA}}^{\w\B}$ [Eq.~(\ref{EFCI+fLDA})] of the 1D He-like atom as a function of the basis size $n_\text{max}$. The first point labelled by ``HF'' corresponds to the limiting case where the basis set $\B$ contains only the exact HF occupied orbital (in this case, FCI simply reduces to HF). The exact energy is taken from Ref.~\onlinecite{MagBur-PRA-04}.}
\label{fig:FCI+fLDA}
\end{figure}

We can now define the \textit{finite} LDA (fLDA) for the complementary multideterminant correlation density functional $\bar{E}_{\c,\md}^{\w\B}[\rho]$ [Eq.~(\ref{EcmdwB})] involved in the second-variant of basis-set correction using the previously determined complementary correlation energy per particle $\bar{\varepsilon}_{\c,\md,\text{fUEG}}^{\w\B}(\rho)$ of the 1D finite UEG
\begin{eqnarray}
\bar{E}_{\c,\md,\text{fLDA}}^{\w\B}[\rho] = \int_{\mathbb{R}} \rho(x) \bar{\varepsilon}^{\w\B}_{\c,\md,\text{fUEG}}(\rho(x)) \d x.
\label{}
\end{eqnarray}
We recall that in standard LDA the functional of an inhomogeneous system for a finite electron number $N$ is approximated using the UEG for infinite electron number $N\to\infty$. Here, instead, in the present finite LDA the functional of the inhomogeneous system is approximated using the UEG of the same electron number $N$. The use of this finite LDA in lieu of the standard LDA should not be seen as a crucial point for the basis-set correction theory but more like a convenient alternative. For a sufficiently short-range complementary interaction $\bar{W}_\ee^\B = W_\ee - W_\ee^\B$, the LDA should not depend much on the electron number used in the definition of the underlying UEG.

We then correct the FCI energy of the 1D He-like atom using this fLDA functional in the non-self-consistent approximation introduced in Eq.~(\ref{EFCI+DFTwB}), obtaining what we will call the ``FCI+fLDA'' energy
\begin{eqnarray}
E_{\text{FCI+fLDA}}^{\w\B} &=& \braket{\Psi_\FCI^\B}{(T + W_\ee + V_\ne)\Psi_\FCI^\B} + \bar{E}^{\w\B}_{\c,\md,\text{fLDA}}[\rho_{\Psi_\FCI^\B}].
\nonumber\\
\label{EFCI+fLDA}
\end{eqnarray}
In practice, we calculate $\bar{E}^{\w\B}_{\c,\md,\text{fLDA}}[\rho_{\Psi_\FCI^\B}]$ by numerical integration using cubic interpolation between calculated values of $\bar{\varepsilon}^{\w\B}_{\c,\md,\text{fUEG}}(\rho)$. The FCI densities of the 1D He-like atom take values from 0 to about 3.5 a.u.. 

In Fig.~\ref{fig:FCI+fLDA}, the FCI+fLDA energy is plotted as a function of the basis size $n_\text{max}$. It is clear that the basis-set correction provides a spectacular improvement of the FCI energy. For example, for $n_\text{max}=0$, the FCI energy is about 55 mhartree above the exact energy while the FCI+fLDA energy is only 1.5 mhartree below the exact energy. For $n_\text{max}\geq 20$, the FCI+fLDA energy is within 1 mhartree of the exact energy. We note that the residual error must come from the fact that in Eq.~(\ref{E0vPsi0wB}) the functional $\bar{E}_{\c,\md}^{\w\B}[\rho]$ is approximated with the fLDA functional and also that the wave function $\Psi_0^{\w\B}$ is approximated by the FCI wave function $\Psi_\FCI^\B$.

\section*{Conclusion}
\label{sec:conclusion}

In this work, we have reexamined the recently introduced DFT-based basis-set correction theory on a 1D model with delta-potential interactions, which is a convenient setting to carefully study the slow basis convergence problem of quantum-chemistry wave-function methods. We provided mathematical details about the formulation of the theory, as well as a new variant of basis-set correction which has the advantage that the basis-set correction functional is defined for all $N$-representable densities. This allowed us to define a LDA for the basis-set correction functional, not based on range-separated DFT as in all previous works, but directly on a 1D finite UEG adapted to the basis set employed. We showed that this approach is very effective to correct for the basis-set incompleteness error in the FCI ground-state energy. We believe that the present work puts the basis-set correction theory on firmer grounds.

Future efforts will focus on the extension of the present work to 3D Coulombic systems. The extension of the theory is straightforward. What remains to be seen is whether the present work adapts well to the standard Gaussian-type-orbital basis sets used in quantum chemistry and whether we can still construct an accurate LDA for the basis-set correction functional based on a 3D UEG with the Coulomb electron-electron interaction projected in the basis set.

\section*{Acknowledgements}
We thank \'Eric Canc\`es, Louis Garrigue, Paola Gori-Giorgi, Muhammad Hassan, Antoine Levitt, Pierre-Francois Loos, and Mi-Song Dupuy for useful comments on the manuscript. We particularly thank \'Eric Canc\`es for pointing out to us the appropriate space ${\cal V}$ of potentials to consider and Antoine Levitt for giving us the correct definition of the space $H_\text{per}^1(\Omega_a^2, \mathbb{C})$. This project has received funding from the European Research Council (ERC) under the European Union's Horizon 2020 research and innovation programme Grant agreement No. 810367 (EMC2).

\section*{Author Declarations}
The authors have no conflicts to disclose.

\section*{Data Availability}
The data that support the findings of this study are available from the corresponding author upon reasonable request.

\appendix
\section{Convergence rate of the expectation value of a Dirac-delta potential in a basis of Hermite functions}

\subsection{One-electron Dirac-delta potential}
\label{app:1e}
Let us consider the 1D hydrogen-like Hamiltonian
\begin{eqnarray}
h = -\frac{1}{2} \frac{\d^2 }{\d x^2} - Z \delta(x),
\end{eqnarray}
with nuclear charge $Z \in (0,+\infty)$. The ground-state wave function is (see, e.g., Refs.~\onlinecite{Gri-BOOK-QM-95,Gel-JAMOP-11})
\begin{eqnarray}
\forall x \in \mathbb{R},\; \varphi(x) = \sqrt{Z} e^{-Z |x|},
\end{eqnarray}
which exhibits a cusp identical to the 3D Coulombic case. We expand $\varphi$ in the orthonormal basis of Hermite functions $\{ f_n^\alpha\}_{n \in \mathbb{N}}$ [Eq.~(\ref{fnx})]
\begin{eqnarray}
\forall x \in \mathbb{R},\; \varphi(x) = \sum_{n=0}^{\infty} c_{n} f_{n}^\alpha (x),
\end{eqnarray}
with coefficients $c_n = \int_{\mathbb{R}} f_n^\alpha(x) \varphi(x) \d x$, which are non-zero only for even integers $n$. Using the following asymptotic equivalent of the unnormalized Hermite functions, for fixed $x$,~\cite{AbrSte-BOOK-83}
\begin{eqnarray}
H_n(\sqrt{2\alpha} x) e^{-\alpha x^2} \isEquivTo{n\to\infty} \frac{2^n}{\sqrt{\pi}} \Gamma\left(\frac{n+1}{2}\right) \cos\left(x \sqrt{4\alpha n} - \frac{n\pi}{2}\right), \;\;\;\;\;
\end{eqnarray}
and the well-known asymptotic equivalent of the gamma function
\begin{eqnarray}
\Gamma(z) \isEquivTo{z\to\infty} \sqrt{2\pi} \; z^{z-1/2} e^{-z},
\end{eqnarray}
we obtain the following asymptotic equivalent of the normalized Hermite functions
\begin{eqnarray}
f_n^\alpha(x) \isEquivTo{n\to\infty} \sqrt{\frac{2}{\pi}} \frac{\alpha^{1/4}}{n^{1/4}} \cos\left(x \sqrt{4\alpha n} - \frac{n\pi}{2}\right). 
\label{fninf}
\end{eqnarray}
Writing $n=2p$ with $p \in \mathbb{N}$, we then find the leading term of the asymptotic expansion of the coefficients $c_{2p}$ by integrating over $x$
\begin{eqnarray}
c_{2p} &\isEquivTo{p\to\infty}& \frac{Z^{3/2}}{\sqrt{2\pi} \; \alpha^{3/4}} \frac{(-1)^p}{(2p)^{5/4}}.
\end{eqnarray}
This is perfectly consistent with the analysis of Refs.~\onlinecite{KvaHjoNil-PRB-07,Kva-PRB-09} which shows that an exponentially decaying function $\varphi$ having a square-integrable first weak derivative (i.e., $\varphi \in H^1(\mathbb{R})$) but a non-square-integrable second weak derivative (i.e., $\varphi \not\in H^2(\mathbb{R})$) must have Hermite expansion coefficients $c_n$ going to zero as $n^{-k}$ with $k\in (1,3/2]$. The leading term of the asymptotic expansion of $c_{2p} f_{2p}^\alpha(x)$ is thus
\begin{eqnarray}
c_{2p} f_{2p}^\alpha(x) \isEquivTo{p\to\infty} \frac{Z^{3/2}}{\pi \sqrt{\alpha}} \frac{1}{(2p)^{3/2}} \cos\left(x\sqrt{8\alpha p}\right),
\end{eqnarray}
and, in particular at $x=0$, 
\begin{eqnarray}
c_{2p} f_{2p}^\alpha(0) \isEquivTo{p\to\infty} \frac{Z^{3/2}}{\pi \sqrt{\alpha}} \frac{1}{(2p)^{3/2}}.
\end{eqnarray}
Calling $\tilde{\varphi}$ the best approximation (in the sense of the $L^2$ norm) to $\varphi$ obtained with a maximal quantum number $n_\text{max}$, i.e.
\begin{eqnarray}
\forall x \in \mathbb{R},\; \tilde{\varphi}(x)&=&\sum_{n=0}^{n_\text{max}} c_{n} f_{n}^\alpha(x),
\end{eqnarray}
we find that $\tilde{\varphi}(0)$ converges slowly to the exact value $\varphi(0)=\sqrt{Z}$ as
\begin{eqnarray}
\tilde{\varphi}(0) &\isEquivTo{n_\text{max}\to\infty}& \varphi(0) - \frac{Z^{3/2}}{\pi \sqrt{\alpha}} \frac{1}{n_\text{max}^{1/2}},
\end{eqnarray}
and the expectation value of the Dirac-delta potential $v_\text{ne}(x)=-Z\delta(x)$ has a similar convergence behavior in $1/n_\text{max}^{1/2}$
\begin{eqnarray}
\braket{\tilde{\varphi}}{v_\text{ne} \tilde{\varphi}} &=& -Z \tilde{\varphi}(0)^2
\nonumber\\
                  &\isEquivTo{n_\text{max}\to\infty}& -Z\varphi(0)^2 + \frac{2Z^{5/2} \varphi(0)}{\pi \sqrt{\alpha}} \frac{1}{n_\text{max}^{1/2}}.
\end{eqnarray}
We also expect the total energy to converge as $1/n_\text{max}^{1/2}$ in a basis of Hermite functions.

\subsection{Two-electron Dirac-delta interaction}
\label{app:2e}

Let us consider the Hamiltonian of the 1D two-electron Hooke's atom~\cite{MagBur-PRA-04}
\begin{eqnarray}
H &=& -\frac{1}{2} \frac{\partial^2}{\partial x_1^2} - \frac{1}{2} \frac{\partial^2}{\partial x_{2}^2} + \frac{1}{2} \omega^2 x_1^2 + \frac{1}{2} \omega^2 x_2^2 + \delta(x_1-x_2), \;\;\;\;\;\;\;
\end{eqnarray}
where $\omega$ is the angular frequency parameter of the external harmonic potential. In contrast to the 1D He-like atom [Eq.~(\ref{H})], the 1D two-electron Hooke's atom has the advantage to be solvable in terms of special functions. Indeed, changing the variables to the center-of-mass (cm) coordinate $X=(x_1 + x_2)/2$ and the relative (rel) coordinate $x_{12} = (x_1 - x_2)$ makes the Hamiltonian separable
\begin{eqnarray}
H &=& h_\text{cm} + h_\text{rel},
\end{eqnarray}
where
\begin{eqnarray}
h_\text{cm} &=& -\frac{1}{4} \frac{\partial^2}{\partial X^2} + \omega^2 X^2,
\end{eqnarray}
and
\begin{eqnarray}
h_\text{rel} &=& - \frac{\partial^2}{\partial x_{12}^2} + \frac{1}{4} \omega^2  x_{12}^2 + \delta(x_{12}).
\end{eqnarray}
The total ground-state energy is then
\begin{eqnarray}
E_0 = {\cal E}_0 + \varepsilon_0,
\end{eqnarray}
where ${\cal E}_0=\omega/2$ is the ground-state energy of $h_\text{cm}$ and $\varepsilon_0$ is the ground-state energy of $h_\text{rel}$ which is found from the equation~\cite{BusHuy-JPB-03,MagBur-PRA-04}
\begin{eqnarray}
2\sqrt{2\omega} \; \frac{\Gamma\left( -\frac{\nu_0}{2} + \frac{1}{2}\right)}{\Gamma\left( -\frac{\nu_0}{2} \right)} = -1,
\end{eqnarray}
where $\nu_0 = \varepsilon_0/\omega-1/2$. For example, for $\omega=1$ a.u., we have $\varepsilon_0 = 0.806746$ a.u.~\cite{MagBur-PRA-04}. The ground-state wave function is
\begin{eqnarray}
\forall (X,x_{12}) \in  \mathbb{R}^2, \; \Psi(X,x_{12}) = \Phi(X) \psi(x_{12}),
\end{eqnarray}
where the center-of-mass wave function is just given by the first Hermite function $\Phi(X) = f_0^{2\omega}(X)$ and the relative wave function is given by~\cite{BusHuy-JPB-03}
\begin{eqnarray}
\forall x_{12} \in \mathbb{R},\; \psi(x_{12}) = c D_{\nu_0}(\sqrt{\omega} |x_{12}|),
\end{eqnarray}
where $c$ is a real-valued normalization constant and $x\mapsto D_\nu(x)$ with $\nu\in \mathbb{R}$ is the parabolic cylinder function~\cite{AbrSte-BOOK-83}. The relative wave function has the same cusp as in Eq.~(\ref{eecusp}), i.e. $\psi(x_{12}) = \psi(0)[1+(1/2)|x_{12}| + O(x_{12}^2)]$.

Let us consider now the expansion of the wave function $\Psi$ in the tensor-product orthonormal basis of Hermite functions $\{(x_1,x_2) \mapsto f_{n_1}^\alpha(x_1) f_{n_2}^\alpha(x_2)\}_{(n_1,n_2)\in\mathbb{N}^2}$. Due to invariance of the harmonic-oscillator Hamiltonian to a rotation of coordinates, the same space is spanned by the rotated orthonormal basis $\{(X,x_{12}) \mapsto f_{n_1}^{2\alpha}(X) f_{n_2}^{\alpha/2}(x_{12})\}_{(n_1,n_2)\in\mathbb{N}^2}$ (see Refs.~\onlinecite{KvaHjoNil-PRB-07,Kva-PRB-09}). This means that the relative wave function $\psi$ is independently expanded as
\begin{eqnarray}
\forall x_{12} \in \mathbb{R},\; \psi(x_{12})&=&\sum_{n=0}^{\infty} d_{n} f_{n}^{\alpha/2}(x_{12}),
\end{eqnarray}
with coefficients $d_n = \int_{\mathbb{R}} f_{n}^{\alpha/2}(x_{12}) \, \psi(x_{12}) \d x_{12}$, which are non-zero only for even integers $n$. Using Eq.~(\ref{fninf}) and with the help of Mathematica~\cite{Math12-PROG-20}, we find the leading term of the asymptotic expansion of the coefficients $d_{2p}$
\begin{eqnarray}
d_{2p} \isEquivTo{p\to\infty} - \frac{(-1)^p \; c}{\alpha^{3/4} 2^{\mu_0 +1/4}\Gamma\left( \mu_0 \right)} \frac{1}{(2p)^{5/4}}.
\end{eqnarray}
where $\mu_0 = - \nu_0/2 +1/2$. Introducing the best approximation to $\psi$ obtained with a maximal quantum number $n_\text{max}$
\begin{eqnarray}
\forall x_{12} \in \mathbb{R},\; \tilde{\psi}(x_{12})&=&\sum_{n=0}^{n_\text{max}} d_{n} f_{n}^{\alpha/2}(x_{12}),
\end{eqnarray}
we find that $\tilde{\psi}(0)$ converges slowly to the exact value $\psi(0)=c \sqrt{2\pi}/[2^{\mu_0}\Gamma(\mu_0)]$ as
\begin{eqnarray}
\tilde{\psi}(0) &\isEquivTo{n_\text{max}\to\infty}& \psi(0) + \frac{c}{\sqrt{\pi \alpha} \, 2^{\mu_0} \Gamma(\mu_0)} \frac{1}{n_\text{max}^{1/2}},
\end{eqnarray}
and the expectation value of the Dirac-delta interaction $W_\text{ee}=\delta(x_{12})$ has a similar convergence behavior in $1/n_\text{max}^{1/2}$
\begin{eqnarray}
\braket{\tilde{\psi}}{W_\text{ee} \tilde{\psi}} &=& \tilde{\psi}(0)^2
\nonumber\\
                  &\isEquivTo{n_\text{max}\to\infty}& \psi(0)^2 + \frac{2c\psi(0)}{\sqrt{\pi \alpha} \, 2^{\mu_0} \Gamma(\mu_0)} \frac{1}{n_\text{max}^{1/2}}.
\end{eqnarray}
We thus see that the two-electron energy converges as $1/n_\text{max}^{1/2}$ in a basis of Hermite functions.


\end{document}